\newcommand{\be}{\begin{equation}}
\newcommand{\ee}{\end{equation}}

\documentclass[journal,10pt]{IEEEtran}

\usepackage{booktabs}
\usepackage{psfrag}
\ifCLASSINFOpdf
   \usepackage[pdftex]{graphicx}

\else
  
   \usepackage[dvips]{graphicx}

\fi
\usepackage[cmex10]{amsmath}
\usepackage{amsthm,epstopdf,nicefrac}

\usepackage{mathtools}

\usepackage[caption=false,font=footnotesize]{subfig}

\newcommand{\tabfinal}{
\begin{table*}\centering
\caption{Properties of the LS--based Estimator of the Square Amplitude of a Sine Wave based on Quantized Samples \label{tabone}}
\ra{1.3}
\begin{tabular}{@{}ll@{}}\toprule
{\bf General properties}   & $\cdot$ asymptotically biased and inconsistent estimator  \\
					&
					$\cdot$ sensitive to errors in the ratio between sampling rate and input sine wave frequency, beyond what can be predicted by the {\em simple} \\ & \hskip2mm {\em approach}   (noise model of quantization)  \\[1.5mm]
{\bf Bias} & $\cdot$ finite for every quantizer resolution ($\Delta$) and any finite value of the number of samples ($N$)\\
	& $\cdot$ not predicted well, when using the {\em simple approach} (noise model of quantization)\\
	& $\cdot$ does not vanish when $N \rightarrow \infty$, for any finite value of $\Delta$\\
	& $\cdot$ reaches quickly -- from a practical engineering viewpoint -- its asymptotic behavior, when $N\rightarrow \infty$\\
	& $\cdot$ spans wider intervals of values, when the signal amplitude increases \\
	& $\cdot$ can be bounded by two expressions showing that the maximum in the magnitude of the bias is approximately dependent on $\Delta^{\frac{4}{3}}$ \\ &   \hskip2mm when $\Delta \rightarrow 0$ \\
	& { $\cdot$ its order of magnitude is insensitive to small offsets in the input signal} \\
	& { $\cdot$ decreases if input additive noise spans at least a quantization step}\\  [1.5mm]
{\bf Variance} 	& $\cdot$ may be both lower and larger than that predicted by the {\em simple approach} (noise model of quantization)\\
	 	& $\cdot$ vanishes quickly --  from a practical engineering viewpoint -- when $N\rightarrow \infty$ \\[1.5mm]
{\bf Mean Square Error} 	& $\cdot$ dominated by the square bias when $\Delta$ is large \\
		& $\cdot$ dominated by the variance when $\Delta$ is small \\
\bottomrule
\end{tabular}
\end{table*}
}

\usepackage{tikz}
\usepackage[bookmarks=false]{hyperref}

\newcommand{\ra}[1]{\renewcommand{\arraystretch}{#1}}
\usepackage{color}

\twocolumn

\newcommand\copyrighttext{%
  \footnotesize \textcopyright 2014 IEEE.  Personal use of this material is permitted. Permission from IEEE must be obtained for all other uses, in any current or future media, including reprinting/republishing this material for advertising or promotional purposes, creating new collective works, for resale or redistribution to servers or lists, or reuse of any copyrighted component of this work in other works. 
 DOI: \href{https://doi.org/10.1109/TIM.2013.2282220}{10.1109/TIM.2013.2282220}
 }
\newcommand\copyrightnotice{%
\begin{tikzpicture}[remember picture,overlay]
\node[anchor=south,yshift=10pt] at (current page.south) {\fbox{\parbox{\dimexpr\textwidth-\fboxsep-\fboxrule\relax}{\copyrighttext}}};
\end{tikzpicture}%
}

\begin{document}
%

\title{A {Rigorous} Analysis of Least Squares Sine Fitting {U}sing Quantized Data: the 
Random Phase Case}

%
%
%

\author{P.~Carbone~\IEEEmembership{Senior Member,~IEEE\thanks{P. Carbone is with the University of Perugia - Department DIEI, via G. Duranti, 93 - 06125 Perugia Italy, on leave at the Vrije Universiteit Brussels, Department ELEC, Pleinlaan 2, B1050 Brussels, Belgium.
}},
and~J.~Schoukens,~\IEEEmembership{Fellow Member,~IEEE}\thanks{J. Schoukens is with the Vrije Universiteit Brussel, Department ELEC, Pleinlaan 2, B1050 Brussels, Belgium.}}%

\newtheorem{theorem}{Theorem}[section]
\newtheorem{lemma}[theorem]{Lemma}

\maketitle

\copyrightnotice

\begin{abstract}
\boldmath
This paper considers least--square based estimation of the { amplitude and} square amplitude of a quantized sine wave,
done by considering random initial record phase.
Using amplitude-- and frequency--domain 
modeling techniques, it is shown that the estimator is inconsistent, biased and has a variance
that may be underestimated if the simple model of quantization is applied. 
{The effects of both sine wave offset values and additive Gaussian noise are taken into account.}
General estimator properties are derived, 
without making simplifying assumptions on the role of the quantization process,
to allow assessment of measurement uncertainty, when this least--square
procedure is used.

\end{abstract}



\newcommand{\fg}[1]{{\frac{1}{\sqrt{2\pi}\sigma} e^{-\frac{{#1}^2}{2\sigma^2}}}} 

%
\IEEEpeerreviewmaketitle

\section{Introduction}
When assessing the properties of 
systems subject to quantized input data, 
the simple noise model of quantization \cite{KollarBook} may be used to obtain results quickly. 
{ However,}
it may also lead to severe approximations when
necessary hypotheses for its application do not hold true. 
This model is based on the assumption that the effects of quantization may be 
represented by a source
of uncorrelated uniformly distributed random variables.
Practitioners can discriminate among applicability situations, 
on the basis of their technical intuition. 
{Nevertheless,}
only sound mathematical modeling can compensate for lack of insight
into the properties of complex algorithms based on quantized data.
{
\subsection{Least Squares Estimation and Quantization}
}
Parametric estimation based on Least Squares (LS), 
is widely used as an all--purpose estimation technique, 
with applications in many engineering domains.
This is the case, for instance, of the
the $3$-- or $4$--parameter sine fit method described in 
\cite{IEEE1241},  used to estimate the meaningful parameters of a sampled sine wave,
{ when testing Analog-to-Digital Converters (ADC)}.
In practice, measurement data 
are almost always obtained by converting analog signals in the digital domain, by means of
an ADC.
As a consequence, they rarely satisfy the conditions for the LS approach to be considered optimal \cite{Schoukens}\cite{Kay}{.
Conversely, the quantization process will result in estimator properties,
that can not easily be analyzed, if the simple noise model of quantization is 
adopted. 
This latter simplification can ease the evaluation of order of magnitudes in 
assessing the estimator bias and variance. 
{ However,} it can be misleading if 
precise measurements and accurate assessments of corresponding uncertainties are necessary. 
This occurs in metrology and in many areas of electrical engineering where quantization plays a role.}
In fact, it is well known that this model breaks down, 
even when very simple input signals enter the processing chain, such as a direct current value.
Nevertheless, engineers often rely on this approximation, and trade accuracy in the analysis of considered systems
for speed and ease in the solution of associated calculations. 
{Also estimations based on a metrological--sound approach can not rely on simplified procedures but
must include all available knowledge needed to provide bounds on estimation errors.}
{ This applies, for instance, when ADCs are tested using sine waves as source signals. The estimation of the 
sine sine wave amplitude needed to estimate other meaningful parameters, e.g. number of effective bits, is done by using
the quantized data at the ADC output \cite{IEEE1241}.}

This work shows that the simple model fails to predict both the bias and the variance
of the LS--based estimator of the amplitude and square amplitude of a quantized sine wave.
{ By adopting several modeling techniques it also}
{{shows how to include the effect of quantization into calculations, without making simplifying assumptions. 
 
The used analytical tools may serve as a fully developed example about how 
solving similar problems, when analogous measuring conditions apply.}}

{
\subsection{The State--of--the--Art}
}

Previous work on the characterization of a LS--based estimator 
applied to the parameters of a sine wave 
subject to Gaussian noise, has been published recently in \cite{Alegria}\cite{Handel}. 
{ Reference \cite{Alegria} contains a detailed analysis
of the LS--based estimator properties, 
when assuming the input sine wave corrupted by additive 
Gaussian noise. It is shown that noise contributes to estimation bias, 
also in absence of quantization.
In \cite{Handel}, the author takes \cite{Alegria} as a starting point and extends its results  
by using alternative derivations and by adding simple closed--form asymptotics, valid for small errors.
This work broadens the analysis to the case in which the sine wave is quantized before
amplitude estimation is performed. 
The same topic is addressed in \cite{KollarBlair} where, 
through simulations, { a description} of the LS--based estimator bias is {obtained}, under the assumption
of quantized data, and a modified technique for bias reduction is illustrated.
Relevant results to this work are presented in \cite{BlairLinnenbrink}
where the definition of the effective number of bits (ENOB) in an ADC is processed to account for 
deviations in testing signal offset and amplitude. Previous work on the same subject was carried 
out in \cite{Pacut2},  where again the definition of ENOB was discussed in relation to the properties of the quantization error characteristics
of a noiseless quantizer and a new analytical method for its assessment was proposed.
} 

\vskip0.5cm
{
\subsection{The Main Results in this Paper}
}

In this paper,
at first a practical example is given to motivate the necessity for this analysis. Then, the 
properties of the estimator of the square amplitude of a quantized sine wave are analyzed, 
under the assumption of randomly distributed initial record phase.
This is a choice of practical interest, because phase synchronization is not always
feasible in the engineering practice, 
when repeated or reproduced measurements results are needed.
{{In these cases, algorithms are applied on quantized data, without knowledge of the initial record phase. Thus, 
their properties can be analyzed by assuming this phase uniformly distributed in the interval $[0, 2\pi)$.}} 
{This assumption is relevant to the analysis of the estimator properties prior to executing measurements. 
When an experiment is performed, a realization of the random variable modeling the initial record phase occurs. From that moment
on, the sequence has a deterministic behavior. Thus, results presented here, serve to understand what to expect before taking 
measurements and how to interpret results when comparing estimations obtained when sampling is not synchronized.}

As it is still customary, to use the noise model of quantization to produce 
estimates of error mean values and variances, the results presented in this paper will 
serve two major purposes: warn against the unconditional application of
{ this model}, as it can lead to underestimates of measurement uncertainties and
show how to use various mathematical tools to avoid simplistic assumptions and to obtain 
estimators with predictable properties.
The described analysis proves that the properties of the 
LS--based estimator of the square amplitude of a quantized sine wave are those listed in Tab.~\ref{tabone}.
The rest of this paper aims at proving these properties: 
while the following sections describe and comment the results in Tab.~\ref{tabone}, 
all mathematical derivations and proofs are put in Appendices,
so to improve readability and usability of results.        
\tabfinal
\vskip0.5cm
\section{System Model and a Motivating Example}
\begin{figure}[h]
\begin{center}
\includegraphics[scale=0.5]{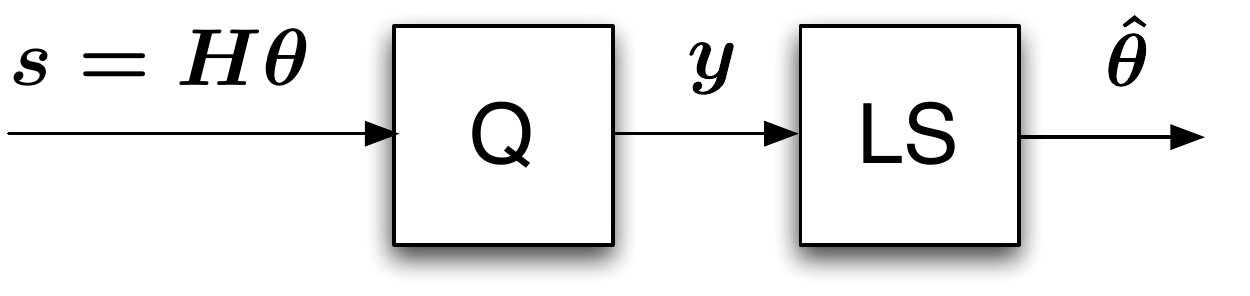}
\caption{Signal chain of the analyzed system. \label{figmodel}}
\end{center}
\end{figure}
\begin{figure}[h]
\begin{center}
\includegraphics[scale=0.45]{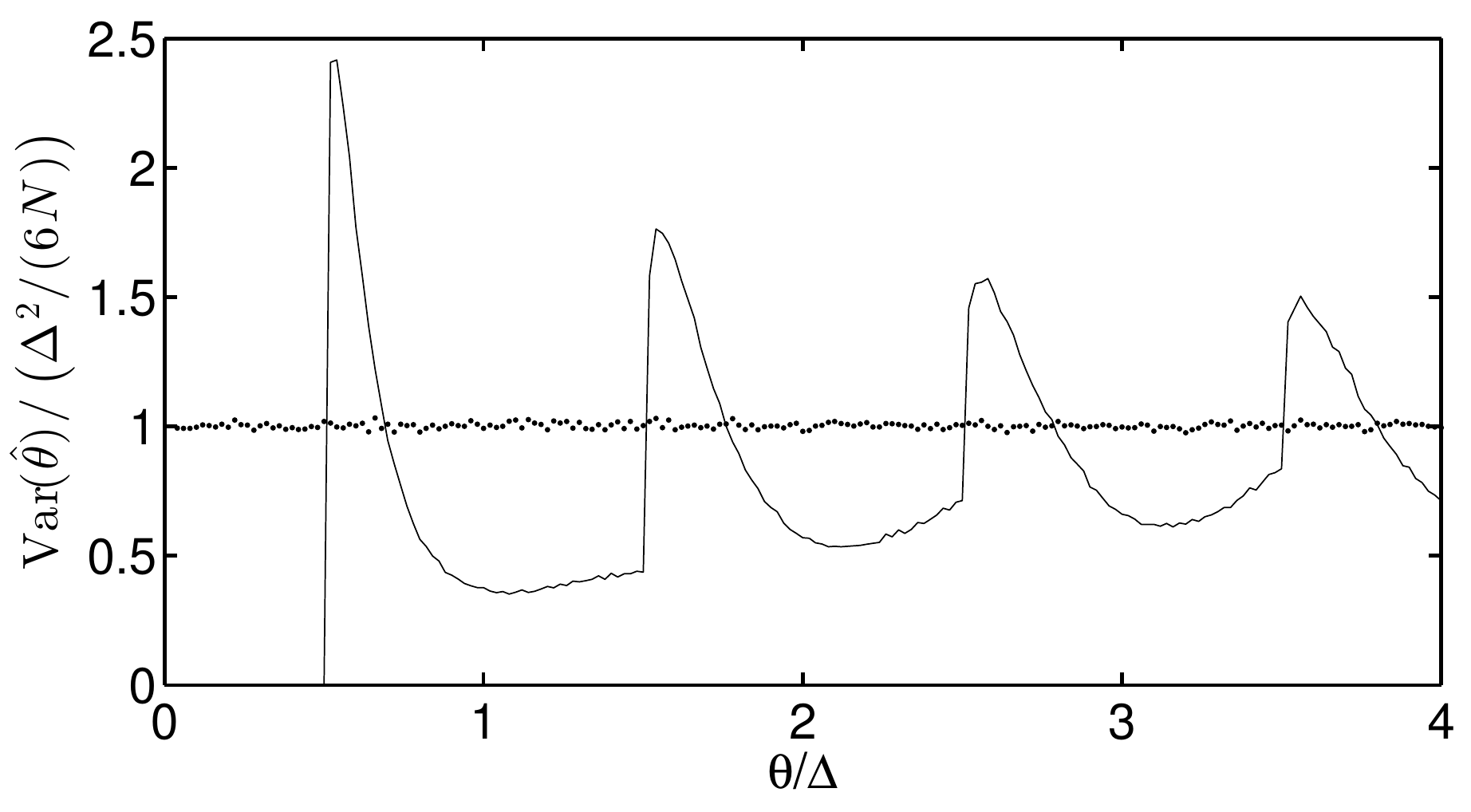}
\caption{Variance of $\hat \theta$ normalized to
the theoretical variance $\Delta^2/(6N)$ obtained by the simple approach ($3$--bit quantizer):
Montecarlo--based estimations based on the
quantizer output sequence (solid line) and on the simplified assumption 
about the error sequence (dotted line). Estimates obtained using a $3$--bit quantizer, $15\cdot 10^3$ records, each one containing $N=200$ samples. \label{figone}}
\end{center}
\end{figure}

\vskip-0.2cm
In this paper, the signal chain depicted in Fig.~\ref{figmodel} is considered.
In this Figure, $s$ represents the input sequence, $\theta$ the vector of parameters to be identified and $H$ 
the observation matrix, that is the matrix with known entries that linearly relates $\theta$ to the input sequence. 
{ This analysis framework is customary in the identification of models that are linear in the parameters {\cite{Kay}}. 
The use of an observation matrix is a more convenient way to express dependencies between parameters and observable quantities,
than using lists of time--varying equations. The specification of the element in this matrix is done on the basis of the 
characteristics of the identification problem to be analyzed and follows standard signal processing approaches \cite{Schoukens}.    
The input sequence} is subjected to quantization, so that the cascaded block performs an LS analysis on a quantized version of the input signal, to provide an estimate $\hat {\theta}$ of $\theta$.
{ To highlight the limits of the simple approach, consider the following example, based on the signal chain depicted in
Fig.~\ref{figmodel}.}
{ Assume ${H}$ as a vector containing} known
samples of a cosine function driven by a uniformly distributed random phase, with independent outcomes
\begin{align*}
&
 {H} \coloneqq [h_0 \cdots h_{N-1}]^T,  
 \quad h_i \coloneqq \cos(\varphi_i) \\
& s_i \coloneqq \theta h_i,  \quad \varphi_i \in {\cal U}(0,2\pi], \qquad i=0, \ldots, N-1,
\end{align*}
and {consider the estimation of the constant amplitude $\theta$}.
By assuming a $3$--bit midtread quantizer, with input range $[-1,1]$, 
the quantizer output sequence $y_i$  
becomes a nonlinear deterministic function of the input $s_i$. The LS--based estimator of $\theta$, 
which also provides the Best Linear Approximation (BLA) of the overall nonlinear system is \cite{Schoukens}
\be
 {\hat \theta} = \left( {H}^T{H}\right)^{-1}  {H}^T{y} = \frac{\frac{1}{N}\sum_{i=0}^{N-1}y_i h_i}{\frac{1}{N}\sum_{i=0}^{N-1}h_i^2}.
\label{esttheta}
\ee
In calculating bias and variance of ${\hat \theta}${{,}} two approaches may be taken.
The first approach ({\em simple calculation}) considers the quantization error as due to a noise source of independent  
random variables with uniform distribution in $\left[ -\frac{\Delta}{2},\frac{\Delta}{2}\right)$
with $\Delta=2/2^b$ as the quantization step and $b$ as the number of quantizer bits.
The second approach ({\em exact calculation}) takes quantization effects into consideration and precisely allows 
assessment of estimator bias and variance. 
It will be shown next how the simple approach may lead to inaccurate results, 
underestimating the effects of quantization. 
By using the simple calculation approach, 
$y_i$ becomes 
\[
y_i = \theta h_i +e_i, \qquad i=0, \ldots, N-1, 
\]
where $e_i$ represents the quantization error, considered independent from $\theta h_i$ and uniformly distributed in
$\left[-\frac{\Delta}{2}, \frac{\Delta}{2}\right)$.
{ Under this assumption,}
in App.~\ref{appeq} it is shown that,
\be
	E(\hat \theta)  \simeq \theta, \qquad \mbox{Var}(\hat \theta) \simeq \frac{\Delta^2}{6N}.
\label{propsimple}
\ee
To verify the validity of the simplified approach, simulations have been carried out.
Both the variance of ${\hat \theta}$ obtained by the simplified calculation and that estimated using the samples at the 
quantizer output are plotted in Fig.~\ref{figone}. { In this Figure, data have been normalized to $\Delta^2/(6N)$ and plotted as a function of the input amplitude normalized to $\Delta$}.
As it is clear from Fig.~\ref{figone}, the simplified calculation provides 
a variance that underestimates the true variance by more than $100\%$ 
for several values of the parameter $\theta$.
This is not surprising because of the strong nonlinear 
distortion introduced by the quantizer, but needs to be addressed.
{ The standard deviation of ${\hat \theta}$ is related to the standard uncertainty in the estimation of $\theta$ {\cite{GUM}}}.
This motivates the analysis of the effects of quantization, 
when LS--based estimators are used to infer parameter values and corresponding uncertainties. 
{ It} is reasonable to assume that by increasing the quantizer resolution, this effect will vanish for large values of $\theta$.
{ At} the same time, it is expected that even in the case of high resolution quantizers this phenomenon will be relevant, especially for small values of the ratio $\theta/\Delta$. 

\section{Least Square Based Estimator of the Square Amplitude of a Quantized sine wave \label{sqa}}
\subsection{Signal and System}
The main estimation problem analyzed in this paper, 
considers an instance of the signal chain depicted in Fig.~1
based on the following assumptions:
\begin{itemize} 
\item (Assumption 1)
The quantizer input signal is a coherently sampled sine wave, defined by
\be
	s_i \coloneqq -A \cos(k_i+\varphi),  \quad k_i \coloneqq \frac{2\pi\lambda}{N}i, \quad i=0, \ldots, N-1 
\label{modeli}
\ee
where $\lambda$ and $N$ are co--prime integer numbers,  $A>0$,
$A^2$ and {$A$ are} the parameters to be estimated and $\varphi$ is the record initial phase.
The coherency condition requires both synchronous sine wave sampling and observation of an integer number of sine wave periods.  
{ Since $s_i$ may also be affected by a deterministic offset or by random noise, subsections~\ref{suboffset}
and ~\ref{subnoise} show the estimator performance under these additional assumptions, respectively.}
\item (Assumption 2)
The variable $\varphi$ is a random variable, uniformly distributed in $[0, 2\pi)$.
\end{itemize}
When $s_i$ is quantized using a mid--tread quantizer, the observed sequence becomes:
\be
	y_i \coloneqq \Delta \left \lfloor  \frac{s_i}{\Delta} +0.5 \right \rfloor= s_i+e_i(s_i), \qquad i=0, \ldots, N-1
\label{midtreadq}
\ee
where $e_i(s_i) \coloneqq y_i-s_i$ represents the quantization error.
As an example, this model applies when testing ADCs \cite{IEEE1241} or in any other 
situation in which the {amplitude} or square amplitude of a sine wave {are} estimated on the basis of quantized data 
(e.g. in the area of power quality measurements). 
Since this is the information usually available when using modern instrumentation, (\ref{modeli}) applies to several situations of interest in the engineering practice.

Assumption 2 is a relevant one because, usually, the value of the initial record phase, 
{{may not be controllable or may be controllable}}
up to a given maximum error. Therefore, when comparing estimates obtained under reproducibility conditions in 
different laboratories or set--up's, the value of $\varphi$ introduces a variability in the estimates. 
{ The consequences} of this variability are the main subject of this paper. 

We will {first} consider the LS--based estimation of $A^2$.
{
This choice  
eases the initial analysis of the estimator properties
since its mathematical expression becomes the summations of weighted products of observable quantities, as shown in (\ref{estima}).
The alternative would be to estimate directly $A$.
However, the additional nonlinear effect of the square--root operation 
implied by extracting $A$ from $A^2$, further complicates the analysis and is treated in 
subsection~\ref{AA}. 
} 
{ According to ({\ref{modeli}}), 
the model to be fitted, by using the observed sequence $y_i$, is
\[
	s_i = -A\cos(\varphi)\cos(k_i)+A\sin(\phi)\sin(k_i) \quad i=0, \ldots, N-1,
\]
that can be rewritten in matrix form as \cite{Kay}:
\[
	S = H\theta
\]
where $S \coloneqq [s_0 \quad s_1 \quad \cdots \quad s_{N-1}]^T$,
\[
{\theta} \coloneqq 
\left[ {\theta}_1 	\quad {\theta}_2  \right]^T \coloneqq
\left[ A\cos(\varphi) \quad A\sin(\varphi) \right]^{T}
\]
is the parameter vector and
\[
H\coloneqq
\left[
\begin{array}{cc}
-\cos(k_0) & \sin(k_0) \\ 
-\cos(k_1) & \sin(k_1) \\
\vdots & \vdots \\
-\cos(k_{N-1}) & \sin(k_{N-1}) 
\end{array}
\right],
\]
is the observation matrix.
Define
\[
{\hat \theta} \coloneqq
\left[ {\hat \theta}_1 	\quad {\hat \theta}_2  \right]^T
\coloneqq 
\left[\widehat{A\cos(\varphi)} \quad  \widehat{A\sin(\varphi)} \right]^{T}
\]
as the estimator of $\theta$.
Since $\theta^2=\theta_1^2+\theta_2^2$, a natural
estimator ${\hat A}^2$ of $A^2$, results in
{{\cite{IEEE1241}}
\be
{\hat A}^2 = {\hat \theta}_1^2 +{\hat \theta}_2^2
\label{ees}
\ee
The LS--based estimator of $\theta$ is \cite{Kay}:
\be
	\hat{\theta} = \left( H^T H\right)^{-1}H^TY
\label{lse}
\ee
where $Y \coloneqq [y_0 \quad \cdots \quad y_{N-1}]^T$.
}
By the coherence hypothesis, and by
the orthogonality of the columns in $H$, 
when $N \ge 3$, 
\[
\left( H^T H\right)^{-1} = \frac{1}{N}\left[
\begin{array}{cc}
2 & 0 \\
0 & 2 \\
\end{array}
 \right]
\]
and (\ref{lse}) provides {\cite{Alegria}}, }  
\[
{\hat \theta}_1 = -\frac{2}{N} \sum_{i=0}^{N-1} y_i \cos\left( k_i  \right),
\qquad
{\hat \theta}_2 = \frac{2}{N} \sum_{i=0}^{N-1} y_i \sin\left( k_i \right) 
\]
that, apart from the signs, are equal to the coefficients of order $\lambda$, in a discrete Fourier transform of the observed sequence.
Thus, from (\ref{ees})
\be
\hat{A}^2 = \frac{4}{N^2}
\sum_{i=0}^{N-1}\sum_{u=0}^{N-1} y_i y_u \cos\left(k_i-k_u \right)
\label{estima}
\ee
Observe that, being a weighted combination of discrete random variables, 
$\hat{A}^2$ is a discrete random variable, whose properties
 are analyzed in the following assuming both 
the simple calculation and the direct calculation { approaches}.

\subsection{Mathematical Modeling \label{matmodel}}
While the subject of quantization has extensively been addressed in the scientific literature, 
it always has proved to be a hard problem to tackle, because of the nonlinear  
behavior of the input--output quantizer characteristic. 
The mathematical modeling used in this paper, is based on several approaches used to
cross--validate obtained results. 
Modeling is performed both in the 
amplitude-- and in the frequency--domains. 
When the Amplitude Domain Approach (ADA) is considered,   the quantizer output is modeled as the sum of
indicator functions weighted by a suitable number of quantization steps $\Delta$, 
extending the same approach presented in \cite{CarbonePetri} (App. \ref{amplitude}).
When the Frequency Domain Approach (FDA) is considered, the quantization error sequence $e_i(\cdot)$ given by
\be
	e_i(s_i) = \frac{\Delta}{2}-\Delta \left \langle \frac{s_i}{2} +\frac{1}{2}\right \rangle, \qquad i=0, \ldots, N-1
\label{qe}
\ee
with $\langle \cdot\rangle$ as the fractional part operator, is expanded using Fourier series, as
done, for example, in \cite{KollarBook} (App. \ref{appbias} and App. \ref{appvar}).   
Both approaches show insights into the problem and
provide closed--form expressions for the analyzed estimator parameters.
Moreover, both  techniques are easy to code on a computer. 
{Thus they can be used to find quick values of error bounds, 
when mathematical expressions
cease to offer easy interpretations of asymptotic behaviors
and of order of magnitudes.}  
A Montecarlo analysis has also been used to confirm mathematical results, indicated in the following as
MA$(R)$, with $R$ being the number of {assumed} records.
In this case, we used both C--language coded LS--based estimators exploiting the LAPACK numerical library, 
to achieve state-of-the-art efficiency in numerical processing and MATLAB, itself a commercial LAPACK wrapper \cite{LAPACK}.  

\subsection{Statistical Assumptions \label{statass}}
(Assumption~3) The {main} system analyzed in this paper is based on the assumption that quantization is applied to noise--free sine wave samples generated by (\ref{modeli}). So would not it be for the random initial phase, the estimator output
would perfectly be predictable, being a deterministic function of system and signal parameters. 
As it is known, wideband noise added before quantization acts as a {\em dithering} signal, that smooths and linearizes 
the quantizer nonlinear characteristics, at the price of an increased estimator's variance \cite{CarbonePetri}. 
Therefore, the {main} analysis described in this paper, provides knowledge about the estimator performance under 
{ the} limiting situations in which the effects of the quantizer finite resolution are dominating over the effects of additive noise.      
This is of practical interest, given that applications exist dealing with the estimation of sine wave amplitudes based on 
very few samples, for low--complexity industrial purposes \cite{Marioli1}\cite{threefive}. Consequently, the assessment of the conditions under which quantization does or does not influence estimators' performance, is necessary to compensate for the lack
of variance suppression benefits usually associated to averaging of a large number of samples.
Because of the Assumption~2, the randomness in the signal is due to $\varphi$, that models the lack of information regarding synchronization between sample rate and sine wave frequency. Therefore, all expected values in the appendices are taken with respect to realizations of $\varphi$.
{When assumption~3 does not apply an additional source of randomness contributes to modify the estimator bias and variance properties. 
This corresponds to the practical situation in which, e.g. when testing ADCs, the input sine wave is corrupted by Gaussian noise.
The analysis of this case is done in subsection~\ref{subnoise}.}

\begin{figure*}[!t]
\centerline{\subfloat{\includegraphics[scale=0.4]{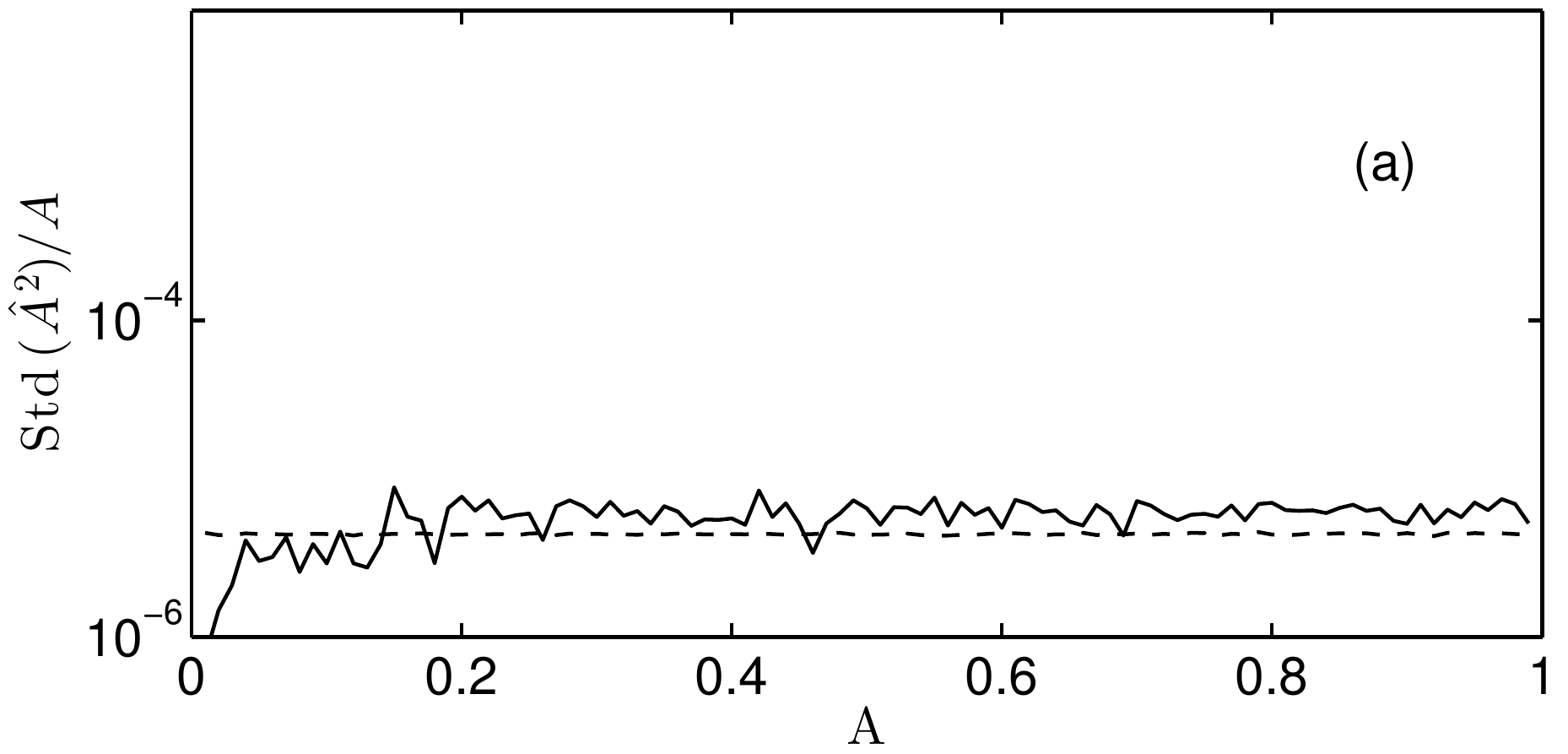}%
}
\hfil
\subfloat{\includegraphics[scale=0.4]{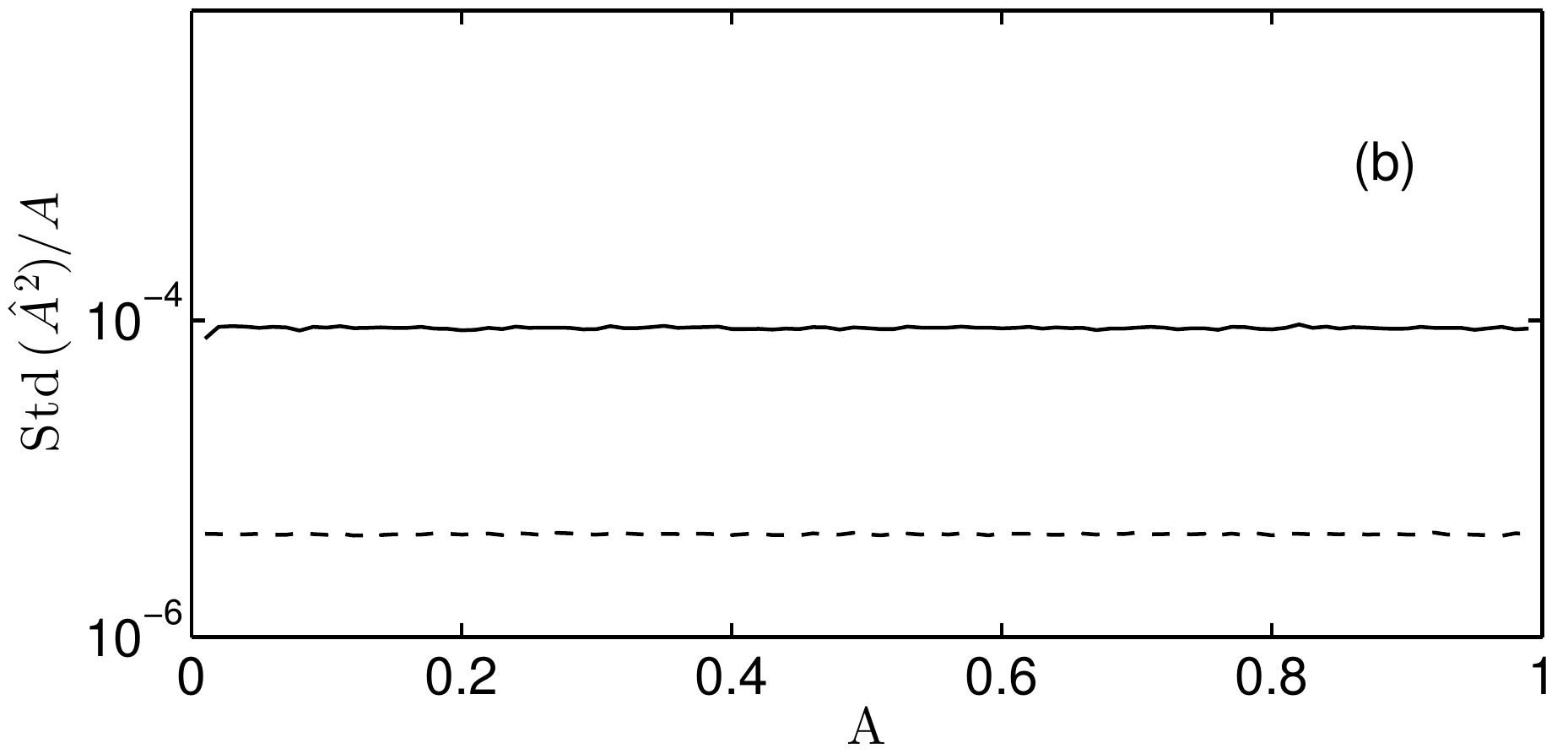}%
}}
\caption{MA$(5\cdot 10^3)$: standard deviation of ${\hat A^2}$ normalized to $A$ when
$N=2\cdot10^3$,  $b=13$: simple approach (dashed line) and
direct approach based on (\ref{estima}) (solid line): (a) $\lambda=201$, (b) $\lambda=200$.}
\label{figtwo}
\end{figure*}

\subsection{Synchronizations issues: Sampling and Sine Fit LS--based Estimation}
As pointed out in \cite{IEEE1241}\cite{Alegria}, 
in practice, the synchronous condition implied by 
Assumption~1, can be met up to 
the frequency errors associated with the equipment used 
to provide experimental data. 
This is the case, for instance, when testing ADCs: 
an accurate sine wave generator provides the test signal to the device under test and 
the quantizer output data are processed to estimate the input signal parameters, needed for additional processing \cite{IEEE1241}. 
In this case, errors in the ratio between the ADC sampling rate and the sine wave frequency may result
in major errors when estimating other related parameters \cite{CarboneChiorboli}. 
When Assumption~1 is not met, also the properties of (\ref{estima}) change significantly. 
As an example, consider the two coprime integers $N=2000$ and $\lambda=201$. 
The normalized standard deviation of ${\hat A^2}$ obtained by the LS--based estimator 
is plotted in Fig.~\ref{figtwo}(a), as a function of $A$ (solid line), when $b=13$.
In the same figure also the normalized estimator standard deviation 
is plotted, when considering the simple approach, that models quantization
as due to the effect of additive uniform noise (dashed line).
Standard deviations in Fig.~\ref{figtwo} are estimated using MA$(5\cdot 10^3)$.
Fig.~\ref{figtwo}(a) shows that the simple assumption almost uniformly underestimates the standard deviation.
Now assume that, due to inaccuracies in the experimental setup, 
the ratio $\lambda/N$ becomes the ratio of two integers with common dividers such as $200/2000=1/10$. 
The same results as those shown in Fig.~\ref{figtwo}(a) are plotted in Fig.~\ref{figtwo}(b). 
Two outcomes are evident: the additive noise model  
still provides the same results as those in Fig.~\ref{figtwo}(a), while that based on
 (\ref{estima}) has a much larger standard deviation,  
increased by $20$ times. This can be explained by observing that{, because of quantization,} only $10$ samples of (\ref{modeli}) having different instantaneous phases are in the dataset. This does not modify the standard deviation in the case of the simple approach because the superimposed 
additive noise still provides useful information, also for those samples associated to the same instantaneous phase. On the contrary,
when noise--free 
data are quantized, samples associated to the same instantaneous phase provide the same quantized output value for a given 
initial record phase $\varphi$ and the estimator uses exactly the same data points affected by strongly correlated quantization error,
over different sine wave periods. 
{In this latter case}, the amount of available information is greatly reduced over the case in which no quantization is applied
and we can conclude that this phenomenon is not modeled properly by the simple approach, 
that provides overly optimistic results. 
When Assumption~1 does not hold true,
and the need is that of also taking into account the finite synchronization capabilities between sample rate and sine wave test frequency, an approach based on the Farey series can be adopted \cite{CarboneChiorboli}.

\begin{figure*}[htbp]
\centerline{\subfloat{\includegraphics[scale=0.4]{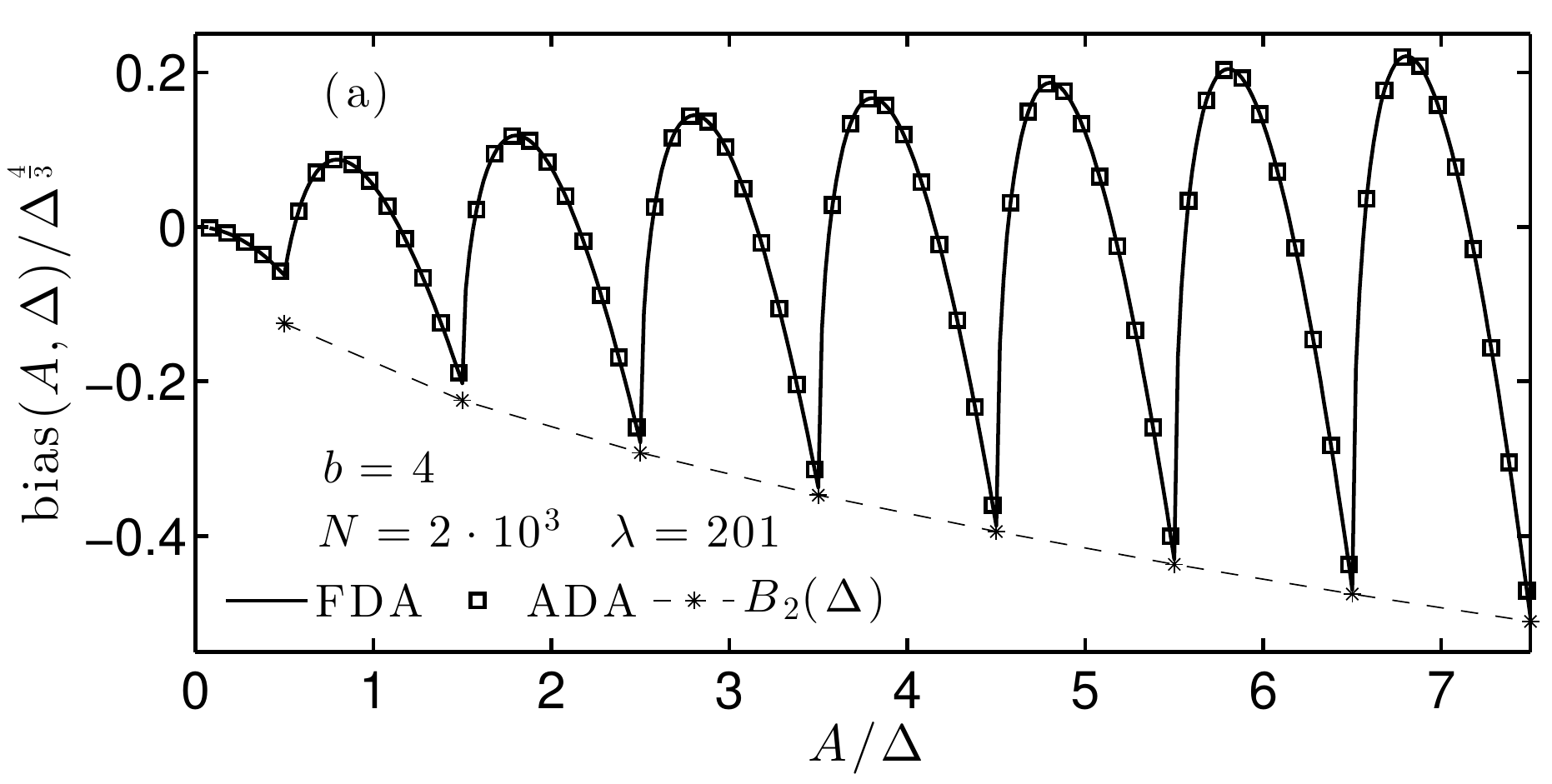}%
}
\hfil
\subfloat{\includegraphics[scale=0.4]{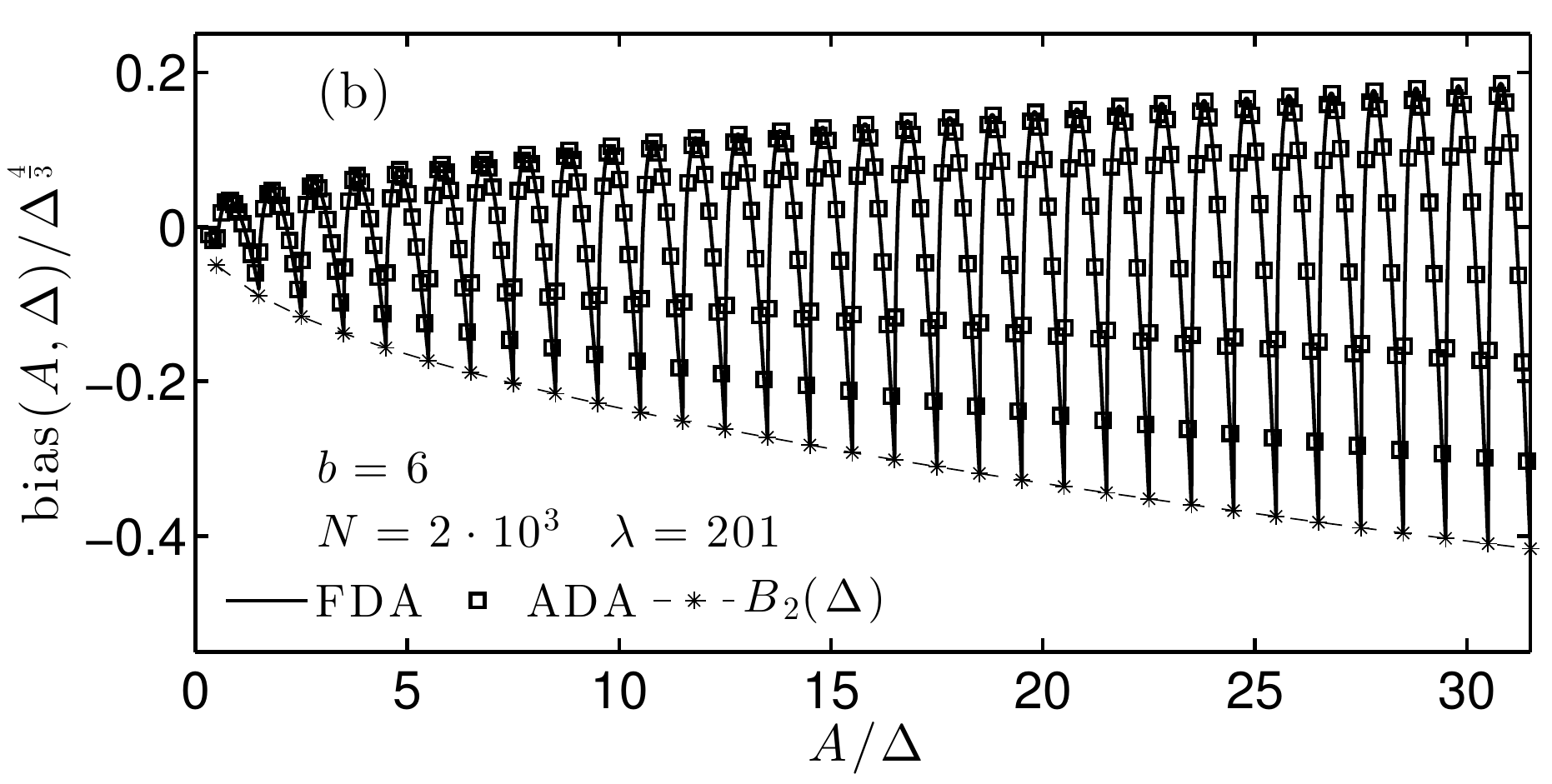}%
}}
\centerline{\subfloat{\includegraphics[scale=0.4]{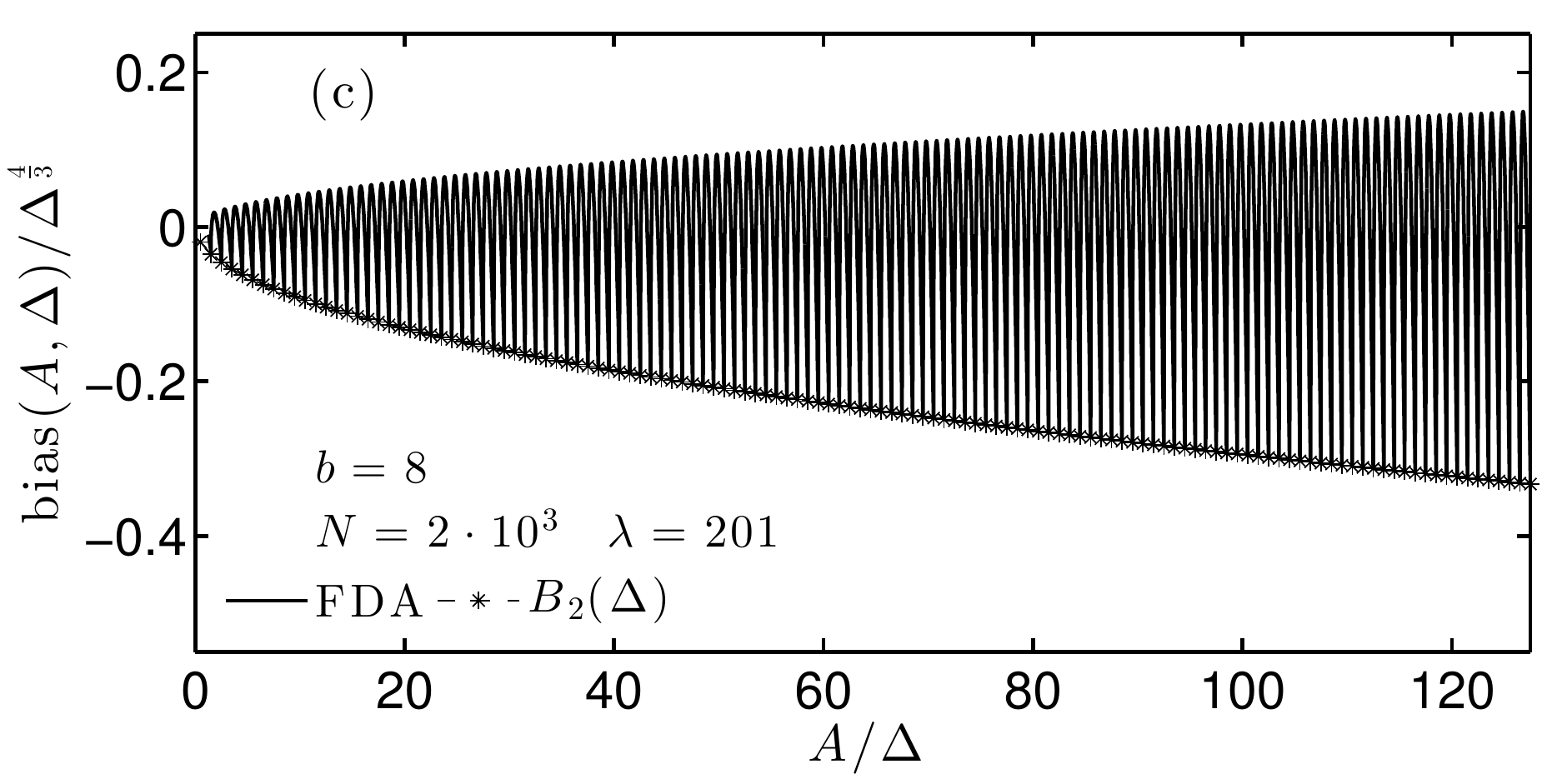}%
}
\hfil
\subfloat{\includegraphics[scale=0.4]{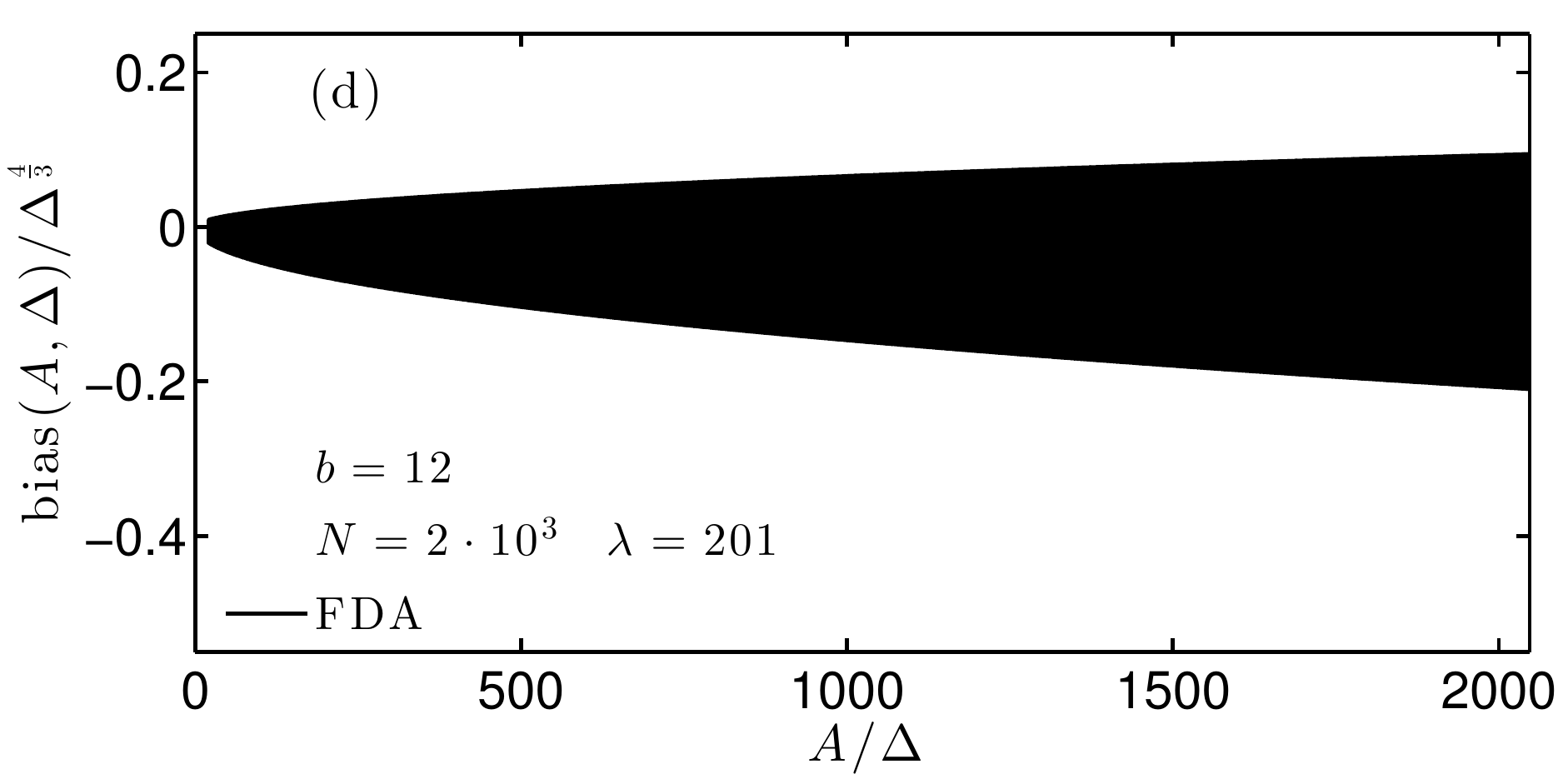}%
}}
\caption{FDA and ADA: normalized bias in the estimation of $A^2$ evaluated by assuming $N=2\cdot10^3$, $\lambda=201$,  $\{4,6,8,12\}$ bit  (FDA--solid line, ADA--squares) and $B_2(\Delta)$ (stars and dashed line).}
\label{figthree}
\end{figure*}

\subsection{Square Amplitude LS--Based Estimation: Bias}
As pointed out in \cite{KollarBlair} the LS--based estimator (\ref{estima}) is biased. 
The bias can be much larger than that predicted by the simple approach.
In App.~\ref{appbias} it is shown that the bias in the estimation of the square amplitude is given by
{
\begin{align}
\begin{split}
	bias(A,\Delta, N) &  \coloneqq E({\hat A}^2)-A^2= \\
	& = 4Ag(A,\Delta)+8h(A,\Delta,N) = \\ 
	& \overset{N \rightarrow \infty }{=}4g(A,\Delta)\left[A+g(A,\Delta)\right] 
	\coloneqq bias(A,\Delta)
\label{biastext}
\end{split}
\end{align}
where
\be
g(A,\Delta) \coloneqq
\left\{
-\frac{x}{2}
+\frac{\sqrt{\pi}}{x\Gamma\left(\frac{3}{2}\right)}
\sum_{k=1}^p\left[ x^2-\left( k-\frac{1}{2} \right)^2\pi^2\right]^\frac{1}{2}
\right\}
\label{ginline}
\ee
with $x\coloneqq \pi A/\Delta$, $\Gamma(\cdot)$ as the gamma function and
\be
p\coloneqq\left \lfloor \frac{x}{\pi}+\frac{1}{2} \right \rfloor=
\left \lfloor \frac{A}{\Delta}+\frac{1}{2} \right \rfloor
\label{pgiven}
\ee
and (App. C),
\begin{multline}
h(A,\Delta,N) \coloneqq
\frac{\Delta^2}{\pi^2 N^2}\sum_{i=0}^{N-1}\sum_{u=0}^{N-1}
\cos(k_i-k_u)  \\
 \cdot
\sum_{k=1}^{\infty}
\sum_{h=1}^{\infty}   
\frac{(-1)^{h+k}}{h k}  
\sum_{n=0}^{\infty}
J_{2n+1}(z_h)
J_{2n+1}(z_k) \\
\cdot 
\cos((2n+1)(k_i-k_u))
\label{hhinline}
\end{multline}
}
Expressions $g(A,\Delta)$ and $bias(A,\Delta,N)$ 
do not uniformly vanish with respect to $A$, for any finite value of $\Delta$, even when $N \rightarrow \infty$.
Thus, (\ref{estima}) is a biased and inconsistent estimator of $A^2$.
Simulations show that few hundreds samples are sufficient for the bias to 
achieve convergence to $bias(A,\Delta)$ when $b < 20$.
Two bounds on $|bias(A,\Delta)|$ 
are derived in App.~\ref{approxg}.
The first one, $B_1(A,\Delta)$, is based on a bound on Bessel functions and is given by:
{
\be
 B_1(A,\Delta) \coloneqq 4AB(A,\Delta)+4B(A,\Delta)^2
\label{B1inline}
\ee
where
\be
B(A,\Delta) \coloneqq \Delta^{\frac{4}{3}} 
\frac{\zeta\left( \frac{4}{3} \right)c}
{\pi\left( 2\pi A\right)^{\frac{1}{3}}},
\label{badelta}
\ee
where $\zeta(s) = \sum_{k=1}^{\infty}\frac{1}{k^s}$ is the Riemann zeta function and
$c=0.7857\ldots$.
}
{
The second one, $B_2(\Delta)$, is based on a finite sum expression of $g(A,\Delta)$ and is given by (see \ref{enve}):
\be
	B_2(\Delta) = 4Ag\left( \left(p-\frac{1}{2} \right)\Delta, \Delta\right)
	\quad p=0, 1, \ldots
	\label{enveinline}
\ee
}
While $B_1(A,\Delta)$ is a (somewhat loose) upper bound on  
$|bias(A,\Delta)|$ for given values of $A$ and $\Delta$, $|B_2(\Delta)|$ 
is a tighter bound, obtained 
with some approximations, that provides the discrete envelope of the minima in $bias(A,\Delta)$ (App.~\ref{approxg}).
The expression for $B_1(A,\Delta)$ shows that  the absolute value of the bias in the estimation of $A^2$  
is on the order of $O\left( \Delta^{\frac{4}{3}}\right)$, 
when $\Delta  \rightarrow 0$.

\begin{center}
\begin{figure*}[t]
\begin{center}
\begin{minipage}{8cm}
\includegraphics[scale=0.4]{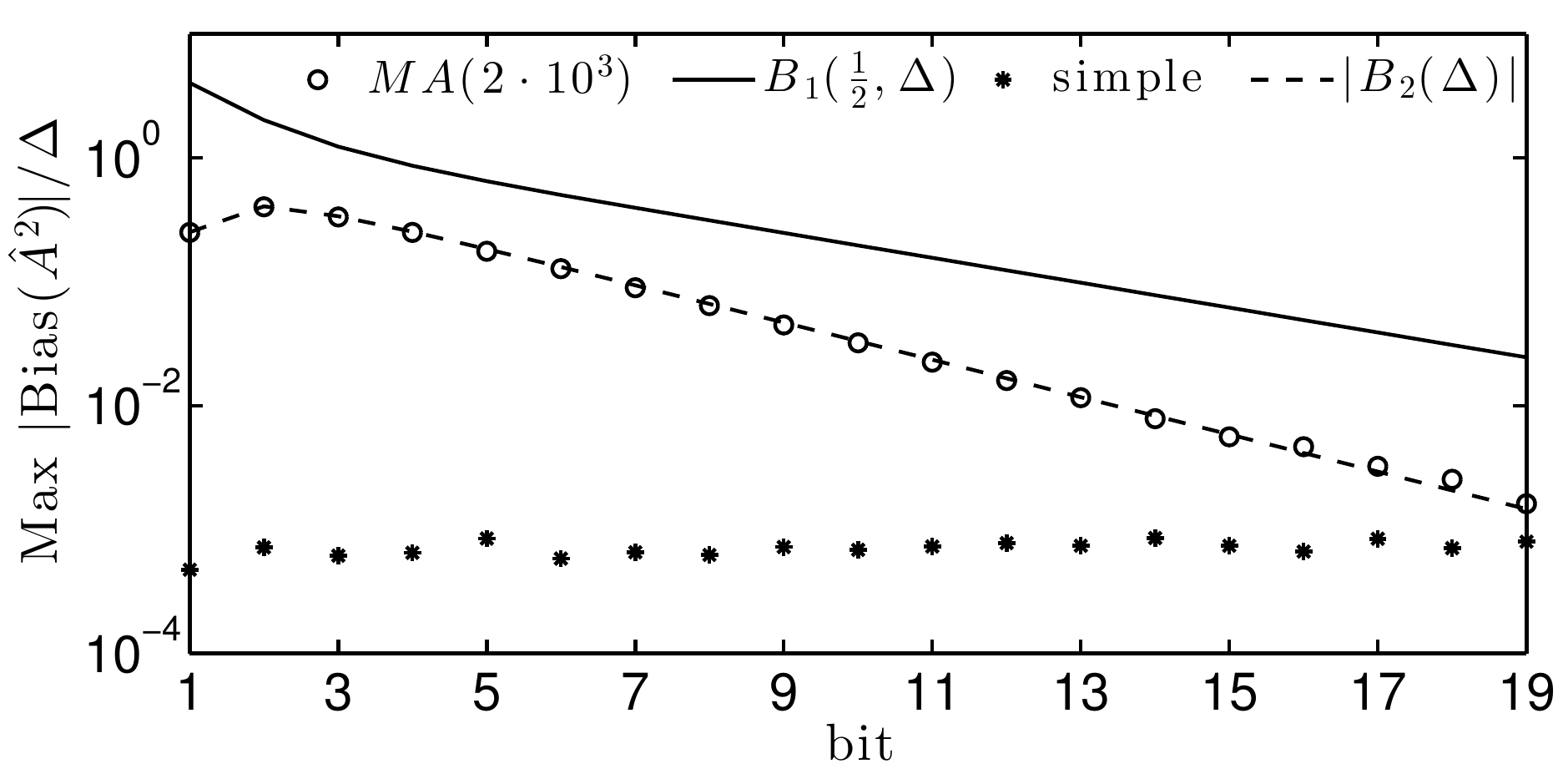}%
\caption{MA$(2\cdot10^3)$ and FDA; normalized maximum in the bias of the square amplitude estimator over all possible values of the input amplitude $0\leq 
A \leq 1-\Delta/2$ (circles) and based on the simple approach (stars).
Shown are also the upper bound $B_1(\frac{1}{2},\Delta)$ (solid line) and the approximate upper 
bound $|B_2(\Delta)|$ (dashed line).  \label{figbias}}
\end{minipage}
\qquad
\begin{minipage}{8cm}
\vskip3.6mm
\includegraphics[scale=0.4]{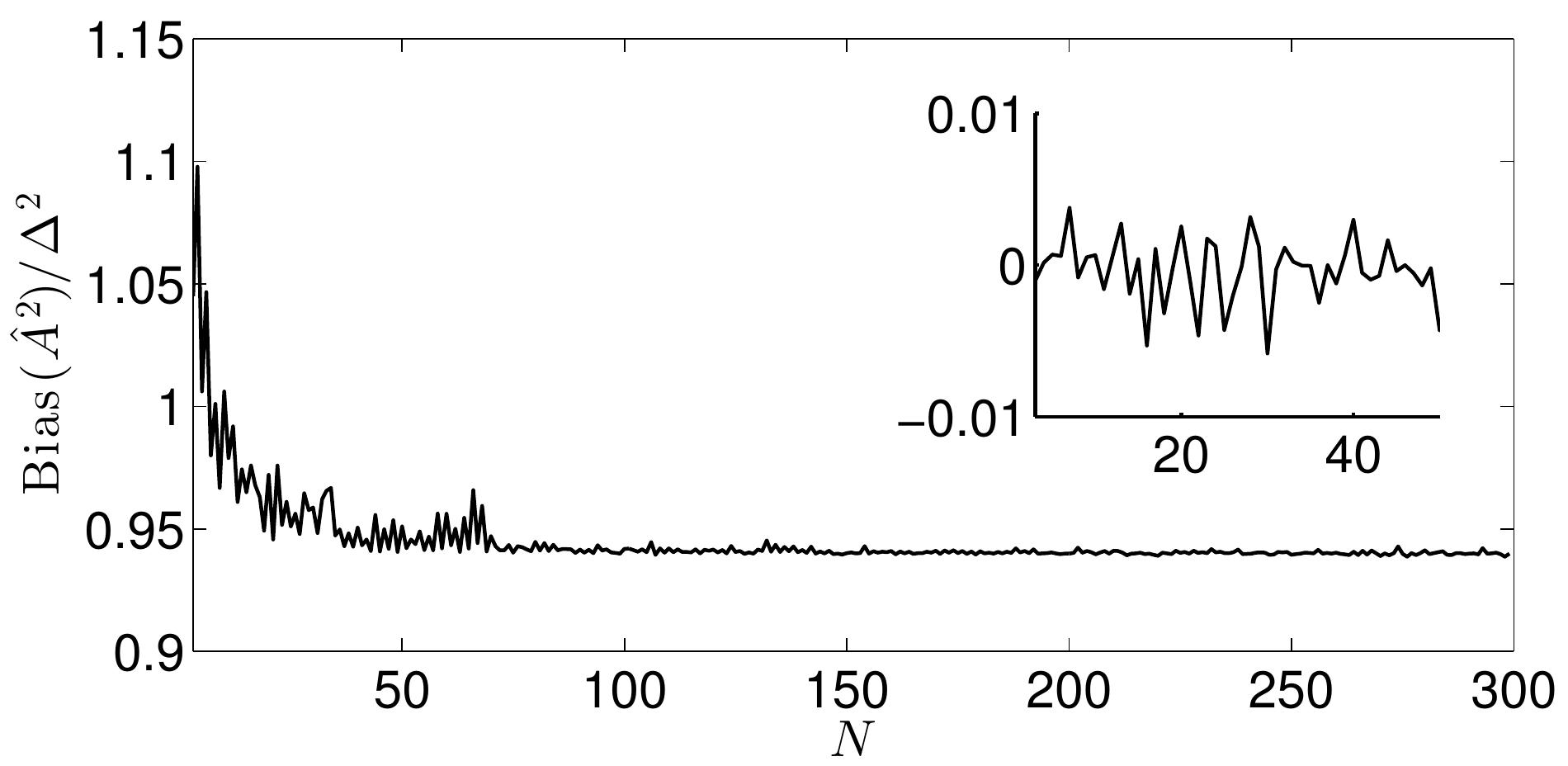}%
\caption{MA$(10^6 \div 5\cdot 10^3)$ and ADA; main figure: normalized bias in the estimation of $A^2$, 
as a function of $N=3, \ldots, 300$, when assuming $b=10$ bit and $A=10.93\Delta$. 
For large values of $N$, the graph converges to $0.9398\ldots$, as predicted by (\ref{schloe}).
In the inset, the difference between the Montecarlo--based estimator and the theoretical expression $bias(A,\Delta,N)$ derived from (\ref{ff}) and normalized to $\Delta^2$. \label{convergebias}}
\end{minipage}
\end{center}
\end{figure*}
\end{center}

To illustrate the behavior of derived expressions, 
$bias(A,\Delta)$
is plotted in Fig.~\ref{figthree}, when assuming $b=4, 6, 8, 12$ bit and $N=2\cdot10^3$.
Plots, calculated using FDA and ADA (solid lines and square symbols, respectively)
have been normalized to $\Delta^{4/3}$ to validate the assumption on the approximate rate
of convergence of the bias with respect to $\Delta$. The approximate equality of the range of values in Fig.~\ref{figthree},
irrespective of the number of bits, supports this statement. 
In Fig.~\ref{figthree}, $B_2(\Delta)$ is plotted using stars joined by a dashed line.
Direct inspection of Fig.~\ref{figthree} shows that the absolute values of 
the minima are larger than those of the maxima. Therefore it is conjectured that  the curve 
obtained through $|B_2(\Delta)|$ is an approximate and tight upper bound on $|bias(A,\Delta)|$.
This hypothesis is confirmed by simulation results based on a MA$(5\cdot10^3)$ and
shown in Fig.~\ref{figbias}. Data are obtained by assuming $N=2 \cdot 10^3$ and $\lambda=539$. 
The maximum magnitude of the bias has been estimated over $0 \leq A<1-\frac{\Delta}{2}$,  when $b=1,\ldots, 19$.
The dashed line represents $|B_2(\Delta)|$ and circles are obtained by estimating the maximum of $|bias(A,\Delta)|$ as a function
of the number of bits.
The maximum of $A$ has been limited to $1-\Delta/2$, so to limit the analysis to granular quantization error,  
when considering quantizers with limited no--overload range, as it happens in practical usage of ADCs.  
The corresponding points obtained by assuming the simple approach are plotted using stars, 
while the continuous line represents the minimum value of the upper bound $B_1(\frac{1}{2},\Delta)$ (App.~\ref{approxg}). 
The plot shows the very good agreement between $|B_2(\Delta)|$, derived through the FDA, 
and maxima obtained using the MA$(5\cdot 10^3)$.

The difference in slopes between $B_1(\frac{1}{2},\Delta)$ and $|B_2(\Delta)|$ explains the
reduction in the range values in Fig.~\ref{figtwo}, as the number of bits increases. 
Consequently, it is conjectured that the absolute value of the bias vanishes faster the $\Delta^{\frac{4}{3}}$ when 
$\Delta \rightarrow \infty$, which is consistent with the large amplitude
approximation of the Bessel functions, for large values of their argument \cite{AbramovitzStegun}. Fig.~\ref{figbias} shows also clearly that the simple approach strongly underestimates the maxima. 
The asymptotic behavior of $bias(A,\Delta,N)$ for large values of $N$ is shown in Fig.~\ref{convergebias}, 
where results based on both the MA$(10^6 \div 5\cdot 10^3)$ and the ADA are shown for a $10$ bit quantizer, when $A=10.93\Delta$. 
Clearly, the convergence rate is quick, as when $N>100$, already the asymptotic value is achieved. 
A variable number of 
records has been used because convergence of Montecarlo--based algorithms depends
on $N$: when $N$ is small a much larger number of records is needed than when $N$ is large.

\begin{center}
\begin{figure*}[htbp]
\begin{center}
\begin{minipage}{8cm}
\includegraphics[scale=0.4]{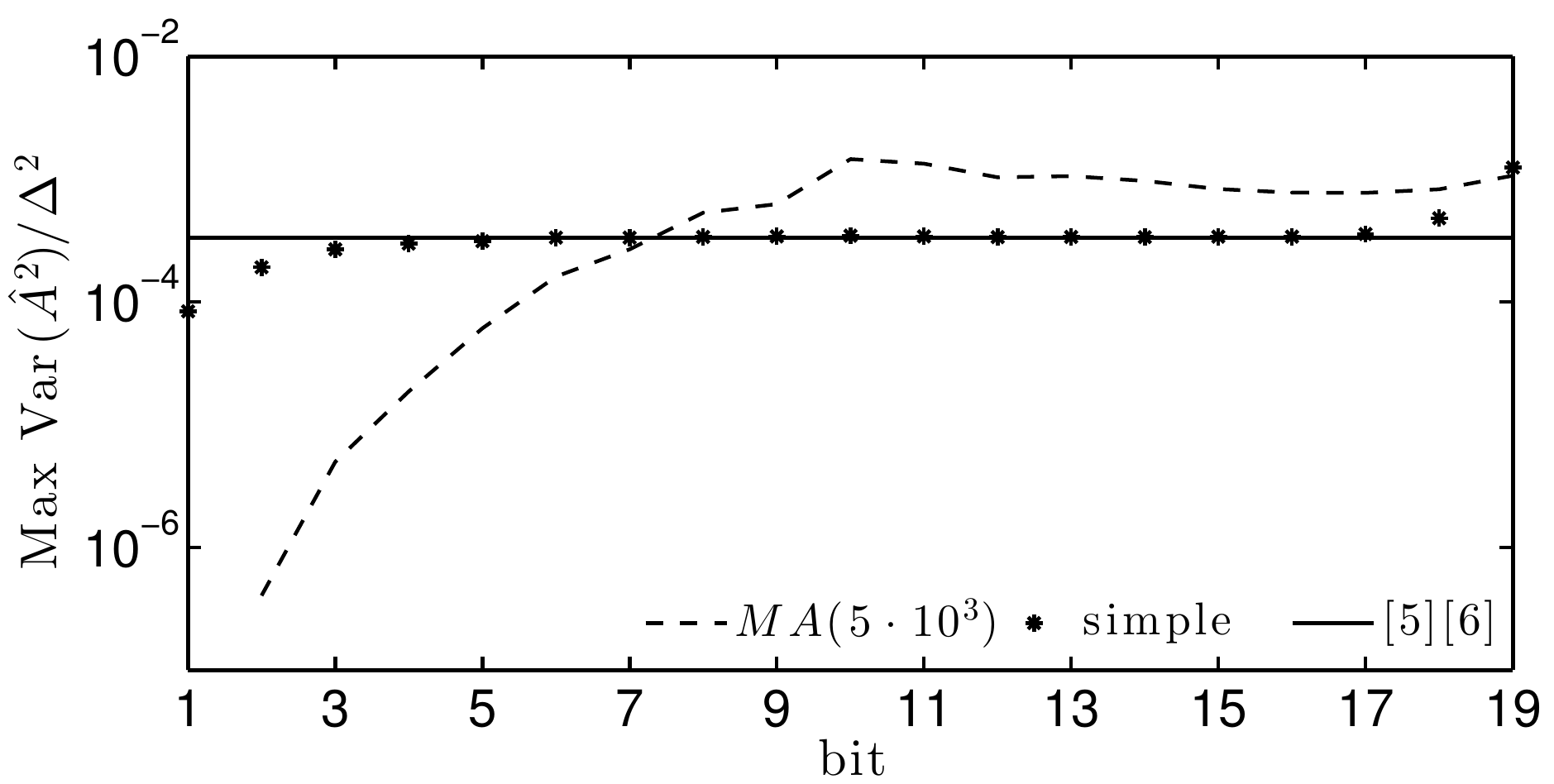}
\caption{MA$(5\cdot 10^3)$, $N=2\cdot 10^3$. Normalized maximum in the variance 
of the square amplitude estimator over all possible values of the input amplitude $0\leq 
A \leq 1-\Delta/2$  (dashed line) and based on the simple approach (dots).
Shown is also the theoretical variance derived in \cite{Alegria}\cite{Handel} under the assumption of zero--mean additive Gaussian noise with variance $\Delta^2/12$ (solid line). \label{mavar} }
\end{minipage}
\qquad
\begin{minipage}{8cm}
\vskip-6.2mm
\includegraphics[scale=0.4]{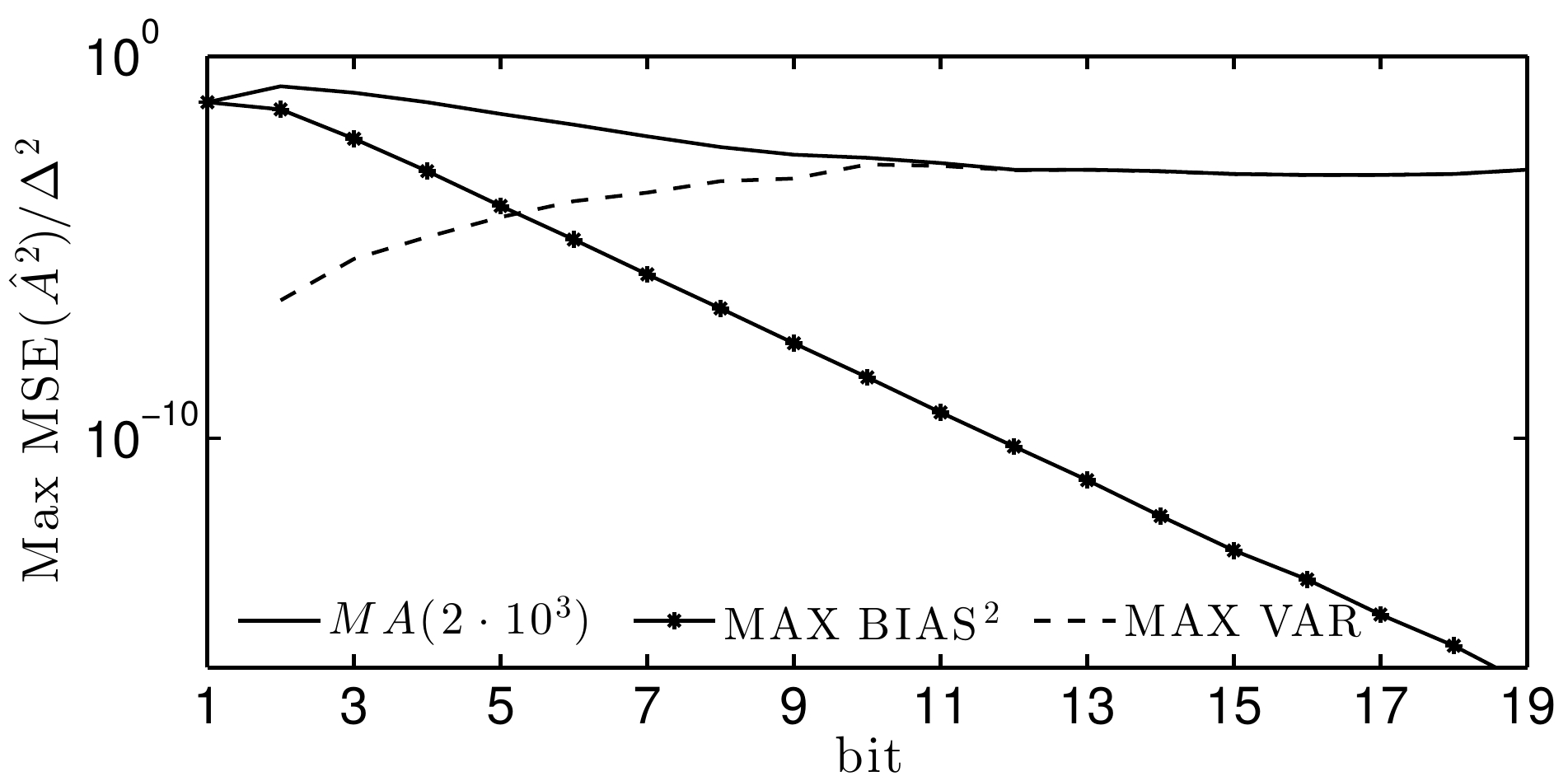}
\caption{MA$(5\cdot 10^3)$ ,  $N=2\cdot 10^3$. Normalized maximum of the mean square error 
over all possible values of the input amplitude $0\leq 
A \leq 1-\Delta/2$  (solid line).
Shown is also the normalized maximum square bias (solid--dotted line)  and variance (dashed line). \label{figmse} }
\end{minipage}
\end{center}
\end{figure*}
\end{center}

\subsection{Square Amplitude LS--Based Estimation: Variance and Mean Square Error \label{lsvar}}
The variance of (\ref{estima}) can again be calculated using the modeling techniques 
listed in Sect.~\ref{matmodel} and applied as in App.~\ref{amplitude} and App.~\ref{appvar}.
To highlight the risks associated to the usage of the simple approach to calculate 
uncertainties in the estimation of the square amplitude, consider the graphs shown in
Fig.~\ref{mavar}. Data in this figure have been obtained using MA$(5 \cdot 10^3)$ under the same conditions
used to generate data in Fig.~\ref{figbias} and represent the normalized maximum
in the variance, over $0<A\leq 1-\frac{\Delta}{2}$: 
the dashed line has been obtained when using quantized data. { Dots} represent the behavior
of the maximum in the estimator variance when the simple model is assumed.
{The} solid line represents its value published in \cite{Alegria}\cite{Handel}, 
obtained {by neglecting quantization and} by {assuming}  
zero--mean additive Gaussian noise, having variance $\Delta^2/12$. 
Clearly, the simple model
underestimates the variance, even when the number of bit is large.
The increasing behavior of the dashed line is due to the large estimator bias  associated with low values
of $b$. When $b$ increases, the bias uniformly vanishes and the variance increases accordingly. 
This is a trend similar to that described in \cite{Handel}, to explain the 
super--efficiency in the behavior of the LS--based estimator under the hypothesis of zero--mean additive Gaussian noise.
This behavior is better seen in Fig.~{\ref{figmse}} where the normalized Mean Square Error (MSE) is plotted together with the 
the normalized square bias and variance, maximized 
over all possible values of $0 < A\leq 1-\frac{\Delta}{2}$,  as
a function of the number of bits.
Data in Fig.~\ref{figmse} are obtained by the MA$(5 \cdot 10^3)$ and considering $N=2\cdot 10^3$ 
sine wave samples, with $\lambda=539$. 
The crossing between curves explains more clearly the super--efficiency type
of effect shown in Fig.~\ref{mavar}. 

\begin{figure}[!ht]
\vskip0mm
\begin{center}
\includegraphics[scale=0.4]{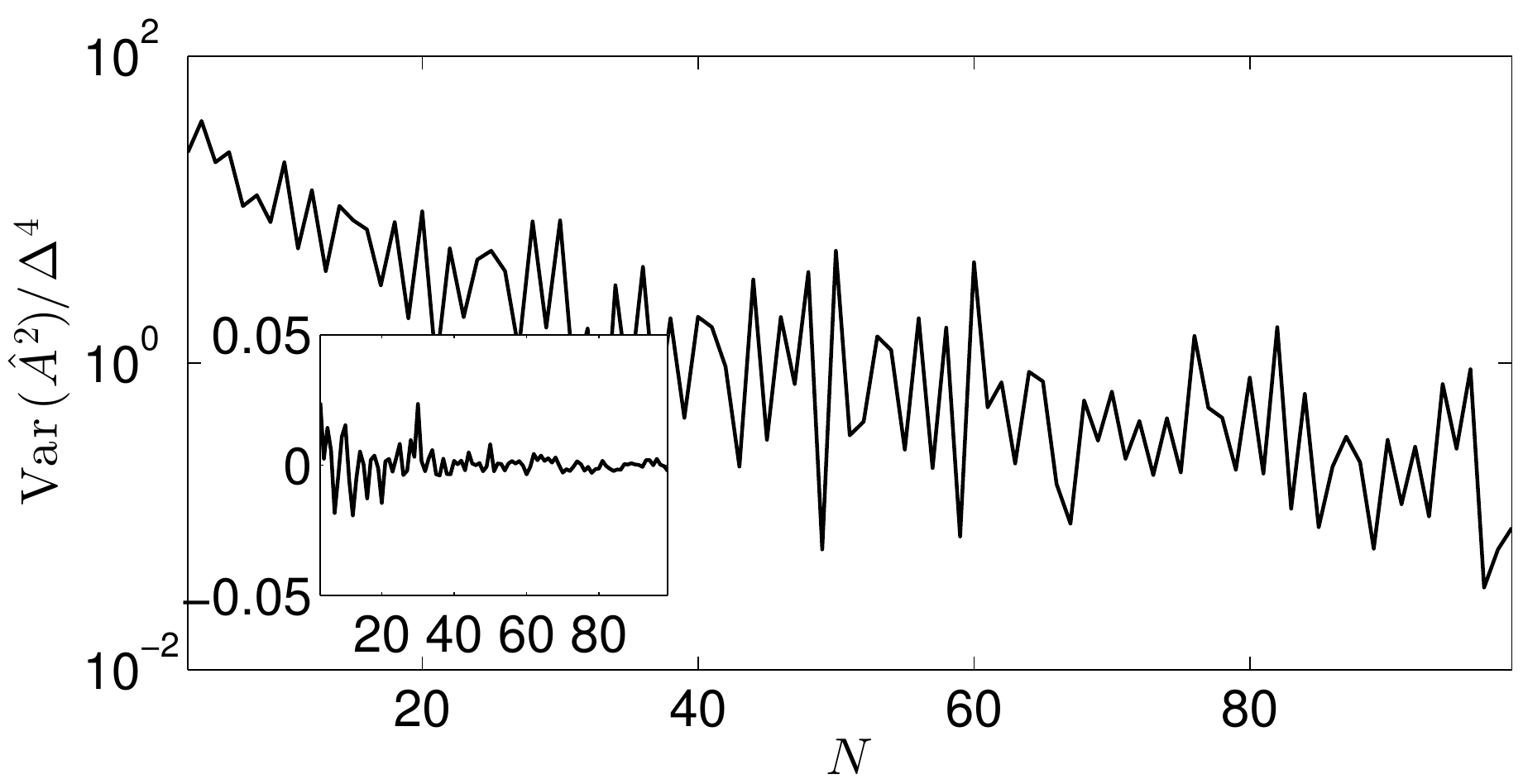}
\caption{MA$(5\cdot 10^3 \div 10^6 )$ and ADA; main figure:
normalized estimator variance as a function of $N$, when $b=10$ and  $A=10.93\Delta$.
Inset: difference between expressions derived using the MA$(5\cdot 10^3)$ and the ADA, normalized to $\Delta^4$. \label{assvar}}
\end{center}
\end{figure}

In App.~\ref{appvar}, it is proved that the variance vanishes when $N \rightarrow \infty$. 
From a practical viewpoint, the speed of convergence
is quick, as shown in Fig.~\ref{assvar}, where the 
MA$(5\cdot 10^3 \div 10^6 )$  has been applied to evaluate the 
estimator variance, assuming $b=10$ bit and $A=10.93 \Delta$. 
A variable number of records has been used in the Montecarlo method,
to keep the simulation time approximately constant for each value of $N$, as $N$ increases.
In fact, when $N$ is small, 
a larger number of records is needed 
to reduce estimators' variance, 
than for larger values of $N$.  

\subsection{
{Amplitude LS--Based Estimation: Bias} \label{AA}}
\begin{figure}[t]
\vskip0mm
\begin{center}
\includegraphics[scale=0.4]{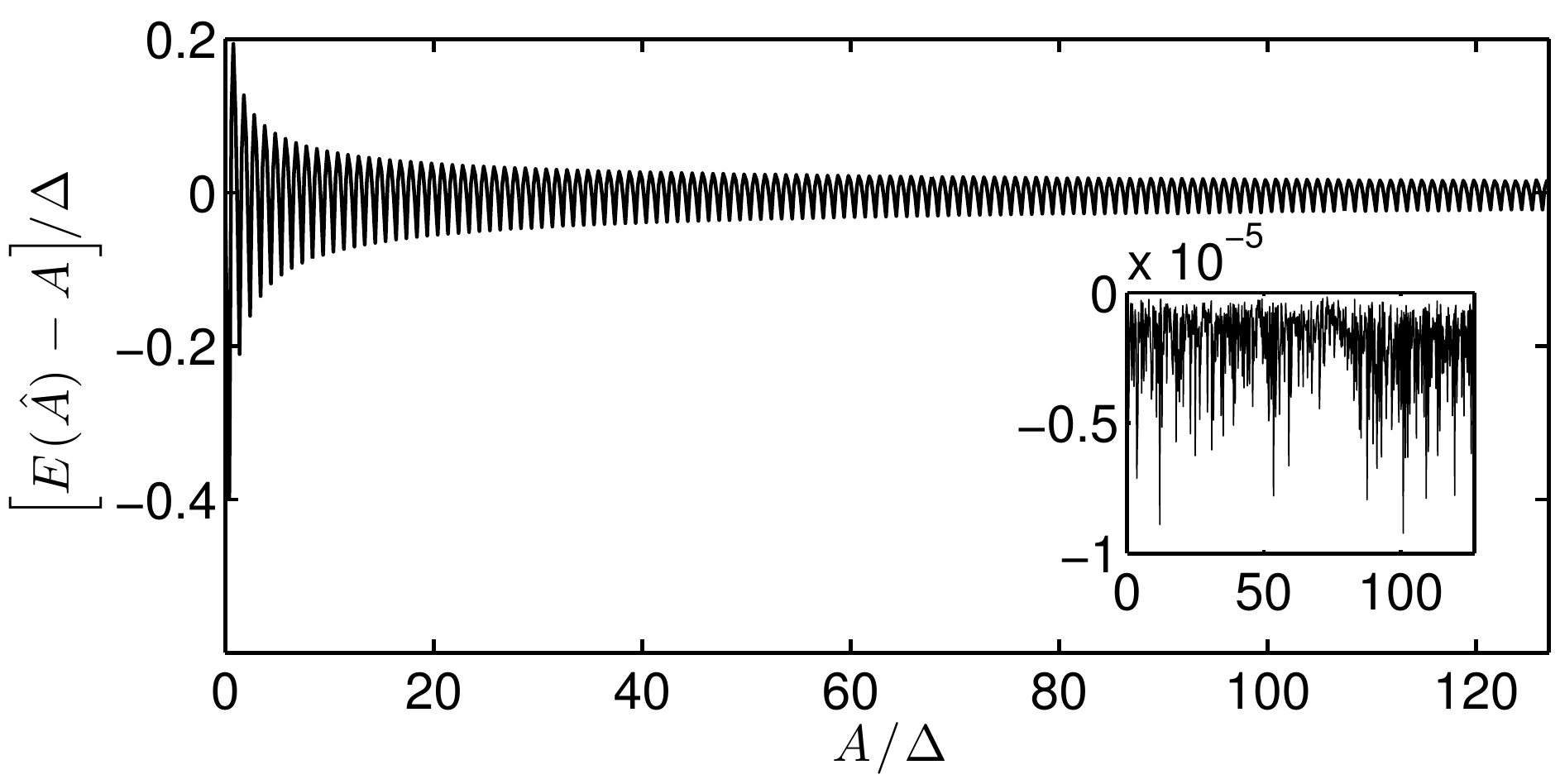}
\caption{MA$(5\cdot 10^3)$ and ADA; main figure:
normalized estimator bias as a function of $A/\Delta$, when $b=10$.
Inset: difference between expressions derived using the MA$(5\cdot 10^3)$ and the ADA, normalized to $\Delta$. \label{figamplitude}}
\end{center}
\end{figure}

{Results} described in section~\ref{matmodel}--\ref{lsvar} allow calculation of the mean value of the LS--based estimator 
${\hat A} \coloneqq \sqrt{\hat A^2}$
of the amplitude $A$ in (\ref{estima}). While, it is well known that nonlinear transformations such as the extraction of the 
square--root, do not commute over the expectation when calculating 
the moments of a random variable, by using a Taylor series expansion about the mean value of ${\hat A^2}$,
we have \cite{Kay}:
\be
	E\left( {\hat A} \right) 
	\simeq 
	\sqrt{E\left( {\hat A^2}\right)} -
	\frac{\mbox{Var}
	({\hat A^2})
	}{8\sqrt{\left[ E\left( {\hat A^2}\right)\right]^3}}	,
	\qquad
	\mbox{Var}
	({\hat A^2}) \rightarrow 0 
	\label{meana}
\ee
When $N \rightarrow \infty$ the variance of the square estimator vanishes as proved in sec.~\ref{lsvar}. 
Therefore, we have 
$	
E\left( {\hat A} \right) 
\overset{N \rightarrow \infty}{\simeq}
\sqrt{E\left( {\hat A^2}\right)}.
$
To illustrate the validity of this approximation, 
results are shown in Fig.~\ref{figamplitude}, in which the normalized bias
in the LS--based estimation of the sine wave amplitude is plotted.
The main figure shows the behavior of the bias,
obtained  by using the MA$(5 \cdot 10^3)$, when
$N=500$, $\lambda=137$ and $b=8$. In the same figure, the inset shows the difference 
in bias estimation when using (\ref{meana}) and the MA($5\cdot 10^3$).
Since the Montecarlo approach does not rely on the approximations induced by the Taylor series expansion,
it proves the validity of this estimation method.
{{
Consequently, all properties 
of the square amplitude estimator
such as the inconsistency, and the behavior of the bias and of its maxima, can easily be adapted by taking the square root 
of the estimators analyzed in { sec.~\ref{matmodel}--\ref{lsvar}},
in the limit $N \rightarrow \infty$.
}}

\begin{figure*}[th]
{\hskip0.6cm
\subfloat{\includegraphics[scale=0.4]{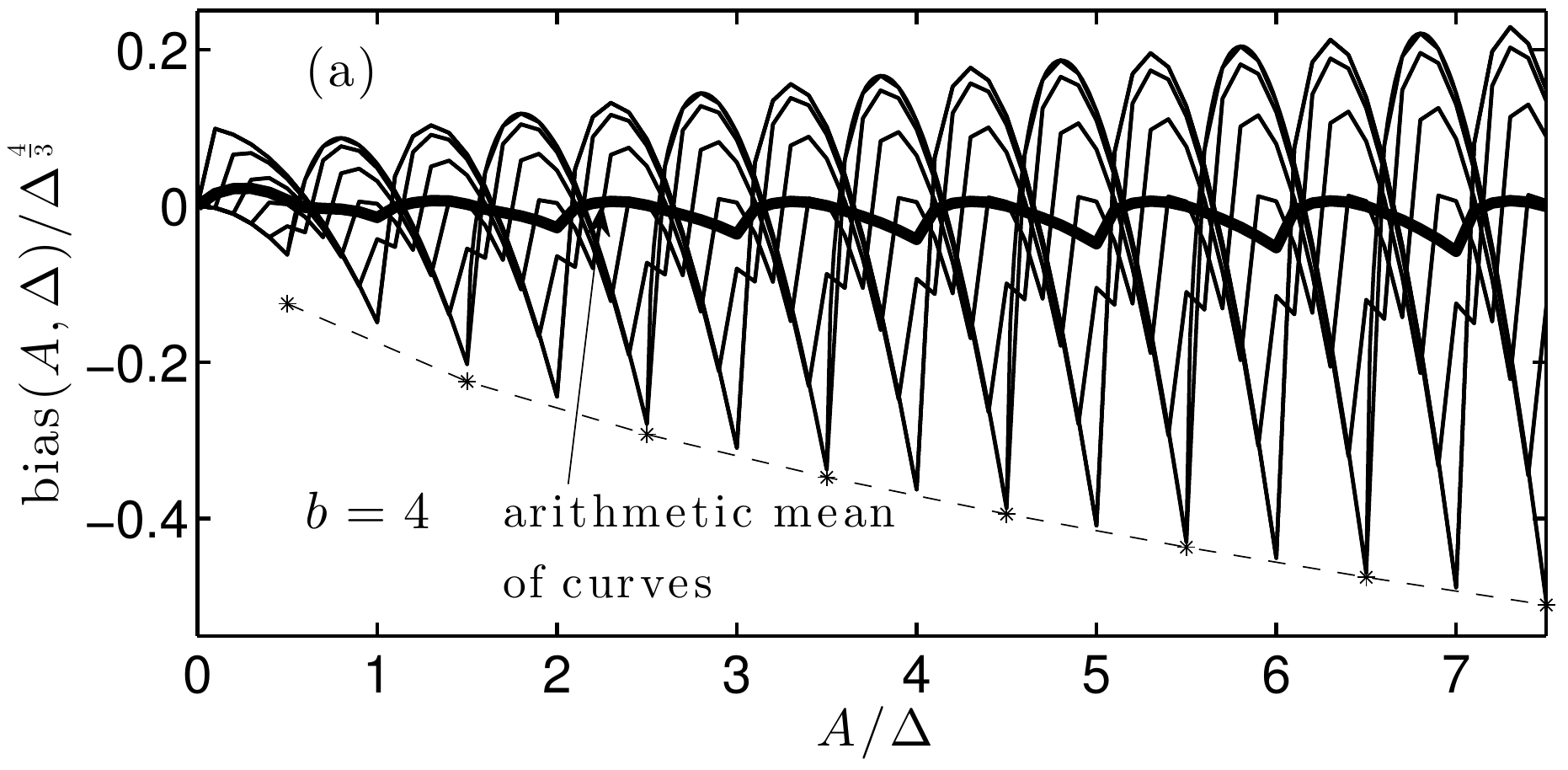}%
}
\hskip3.2cm
\subfloat{
\hskip-2cm\includegraphics[scale=0.4]{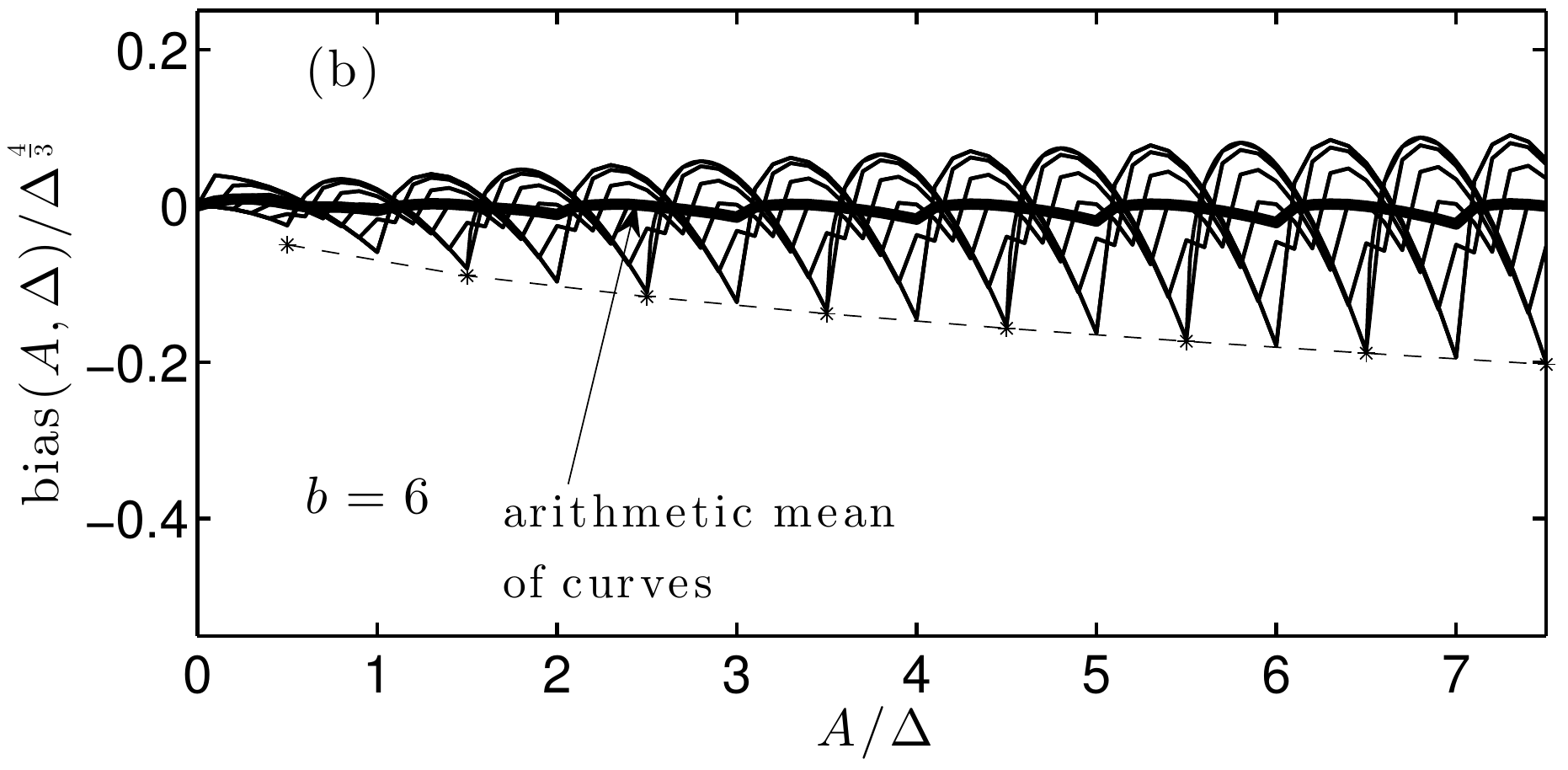}%
}}
\caption{{ ADA: normalized bias in the estimation of $A^2$ evaluated by assuming $N=2\cdot10^3$, $\lambda=201$,  $\{4,6\}$ bit
assuming an input sine wave offset taking values in $\{-\nicefrac{\Delta}{2}, -\nicefrac{\Delta}{2}+\nicefrac{\Delta}{10}, 
-\nicefrac{\Delta}{2}+\nicefrac{2\Delta}{10}, \ldots, \nicefrac{\Delta}{2} \}$.
Shown is also the arithmetic mean of the bias curves and $B_2(\Delta)$ (stars and dashed line).}}
\label{figtbiasoffset}
\end{figure*}

\subsection{
{Squared Amplitude LS--Based Estimation Bias: Effect of Input Offset} \label{suboffset}}
{
The models described in this paper allow the analysis of the estimator performance also when the input
signal has a non--zero mean value, that is when (\ref{modeli}) is modified to include an additive constant
$d$ as follows:
\[
s_i \coloneqq -A\cos(k_i+\varphi)+d, \quad k_i=\frac{2\pi \lambda}{N}i, \quad i=0,\ldots, N-1.
\]
As an example, this case is relevant when the sine wave is used as a test signal for assessing ADC performance.
Precisely controlling the mean value up to the accuracy required by the quantization step width, may not 
be an easy task, especially for small ADC resolutions. At the same time, temperature and voltage 
drifts may induce variations in input--related offset voltages in the considered ADC. 
Therefore, bounds on the estimation error need to include also the effect of $d$.
Including an offset
in the input signal may be equivalent to modeling a quantizer other than the mid--tread one considered in (\ref{midtreadq}), acting
on an offset--free sine wave. As an example, if $d=-\left( \nicefrac{\Delta}{2}\right)$, (\ref{midtreadq}) becomes the model of a truncation quantizer applied to (\ref{modeli}). To verify the implications of an additive constant in the input signal, the bias in estimating 
$A^2$ has been evaluated using the $ADA$, under the same conditions used to generate data shown in Fig.~\ref{figthree}.
In Fig.~\ref{figtbiasoffset}(a)--(b), corresponding to Fig.~\ref{figthree}(a)--(b), is shown the normalized bias,  when assuming 
$b=4, 6$. In this Figure, $d$ is assumed as a parameter taking values in $-\left(\nicefrac{\Delta}{2}\right), \ldots, \nicefrac{\Delta}{2}$ in 
steps of $\nicefrac{\Delta}{10}$. The overall behavior shows that the added offset does not worsen the bias with respect to data shown 
in Fig.~\ref{figthree}. The bold line in Fig.~\ref{figtbiasoffset} is the arithmetic mean of all curves. Thus, it approximates 
the behavior of the bias when $d$ is taken as being a random variable uniform in $\left[ -\frac{\Delta}{2},  \frac{\Delta}{2}\right)$. }

\begin{center}
\begin{figure*}[hbpt]
\begin{center}
\begin{minipage}{8cm}
\includegraphics[scale=0.4]{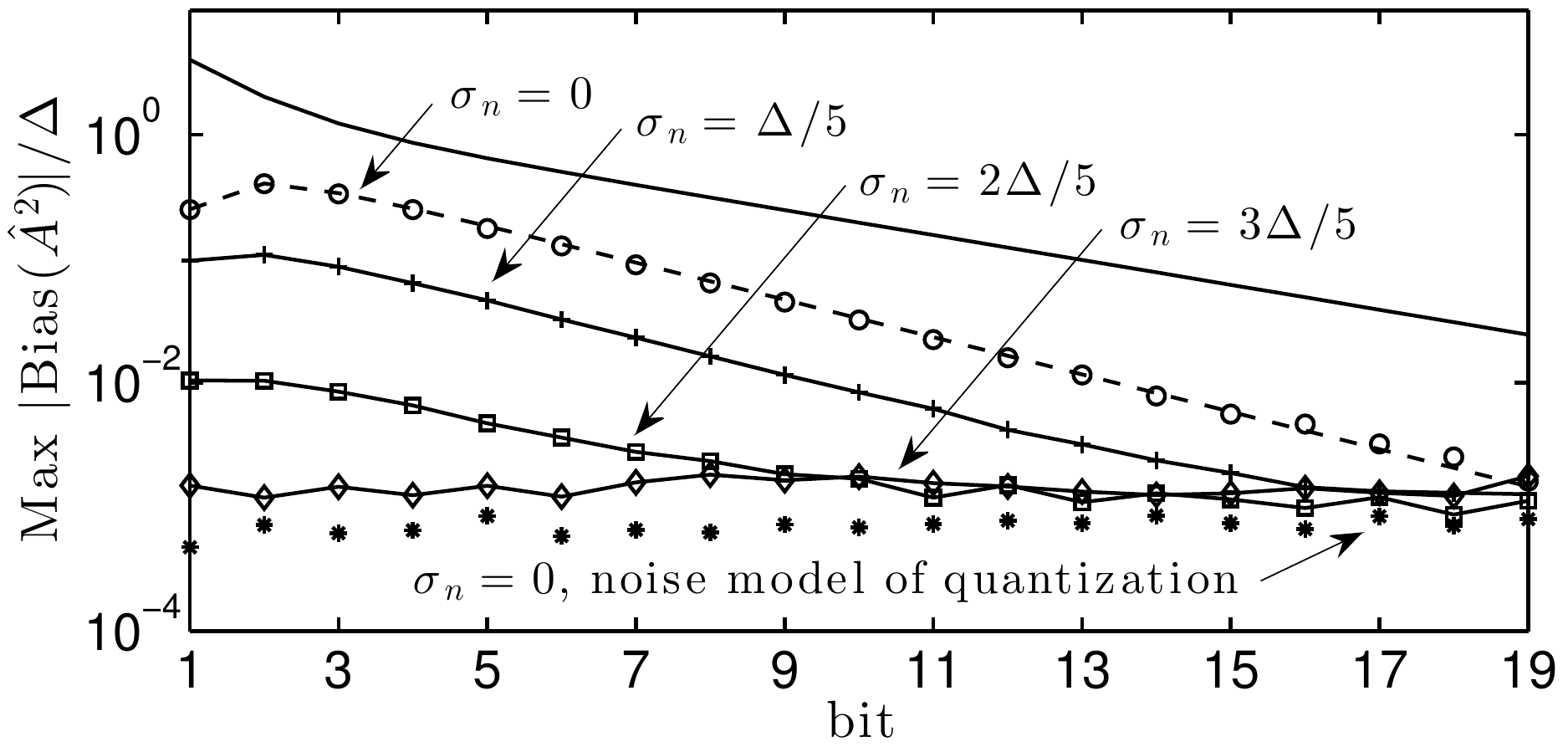}
\caption{{ MA$(2\cdot10^3)$: normalized maximum in the bias of the square amplitude estimator over all possible values of the input amplitude $0\leq 
A \leq 1-\Delta/2$ (circles), by assuming zero--mean additive Gaussian noise having 
the indicated standard deviation. Shown is also  
the normalized maximum obtained by the simple approach (stars),
the upper bound $B_1(\frac{1}{2},\Delta)$ (solid line) and the approximate upper 
bound $|B_2(\Delta)|$ (dashed line).} \label{fignoise_bias} }
\end{minipage}
\qquad
\begin{minipage}{8cm}
\includegraphics[scale=0.4]{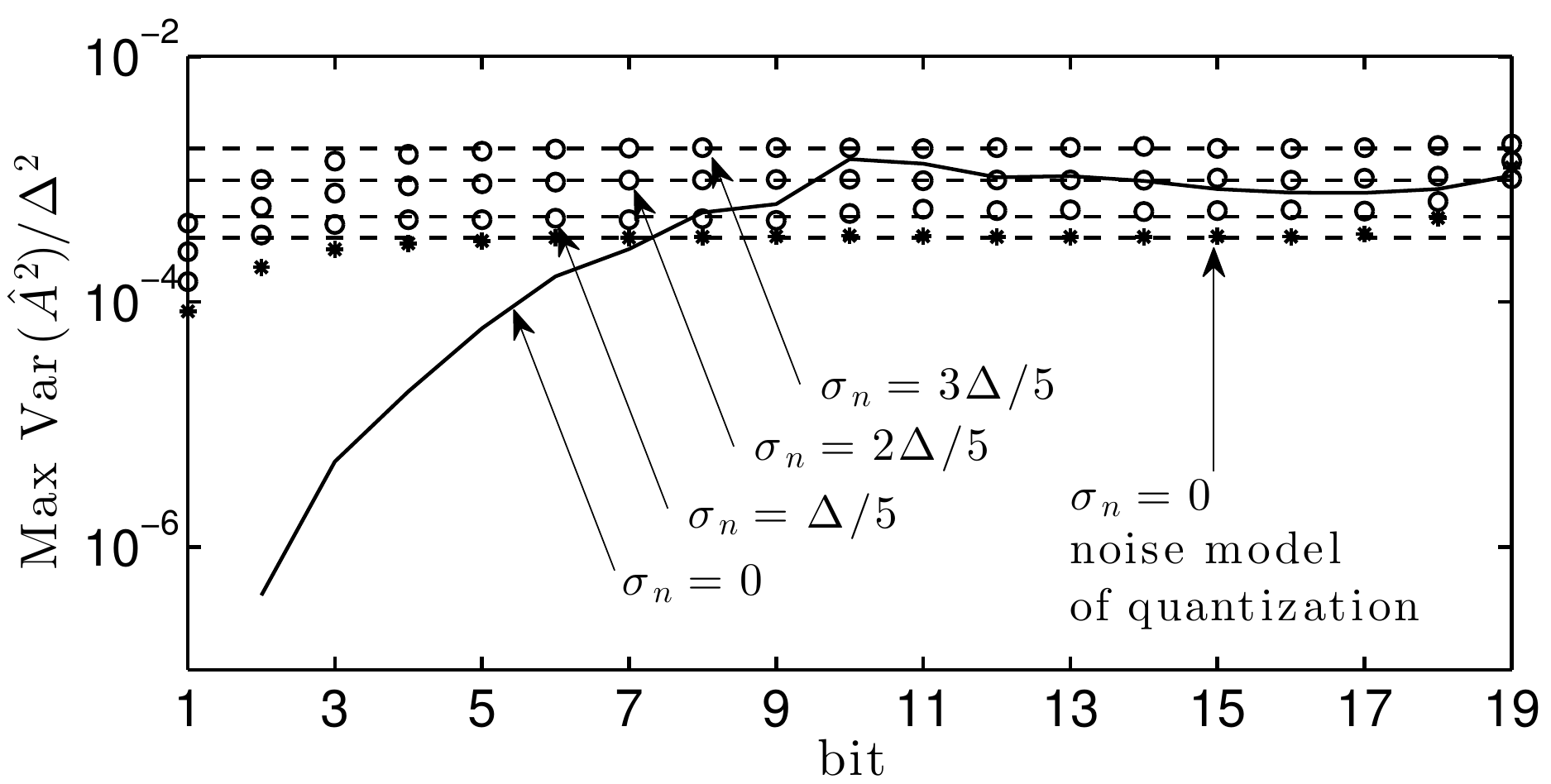}
\caption{{ MA$(5\cdot 10^3)$, $N=2\cdot 10^3$. Normalized maximum in the variance 
of the square amplitude estimator over all possible values of the input amplitude $0\leq 
A \leq 1-\Delta/2$  (circles and solid line), by assuming zero--mean additive Gaussian noise having 
the indicated standard deviation.
Shown is also the normalized maximum obtained by 
the simple approach (stars) and the theoretical variance derived in \cite{Alegria}\cite{Handel} under the assumption of zero--mean additive Gaussian noise with variance $\Delta^2/12$ (dashed lines).}  \label{fignoise_var} }
\end{minipage}
\end{center}
\end{figure*}
\end{center}

\subsection{
{Squared Amplitude LS--Based Estimation Bias: Effect of Input Noise\label{subnoise}}}
{
The quantizer input sine wave may be affected by additive wide--band noise $n_G$, as follows:
\[
s_i \coloneqq -A\cos(k_i+\varphi)+n_G, \quad k_i=\frac{2\pi \lambda}{N}i, \quad i=0,\ldots, N-1.
\]
As it is well--known, noise act as a {\em dither} signals ({\em unsubtractive} dither in this case) and under suitable conditions, it renders the quantization
error a uniformly distributed random variable, regardless of the input signal distribution \cite{KollarBook,WannamakerLipshitz}.
The overall effect of properly behaving input--referred 
additive noise is that of linearizing the 
mean value of the quantizer input--output characteristic. Consequently, a reduction in the estimation bias shown in Fig.~\ref{figthree} and
Fig.~\ref{figtbiasoffset} is expected. This will come at the price of an increased estimation variance. In fact, an additional source of uncertainty
characterizes the estimator input data, beyond that taking into account the initial record phase.
Additive zero--mean Gaussian noise has been considered in the following, having standard deviation $\sigma_n \in \{0, \nicefrac{\Delta}{5}, \nicefrac{2\Delta}{5}, \nicefrac{3\Delta}{5} \}$. 
To verify the implications of additive Gaussian noise, the bias in estimating 
$A^2$ has been evaluated using the $MA(2\cdot 10^3)$, under the same conditions used to generate data shown in Fig.~\ref{figbias} and 
Fig.~\ref{mavar}. Results are plotted in Fig.~\ref{fignoise_bias} and Fig.~\ref{fignoise_var}, respectively.
Plots in Fig.~\ref{fignoise_bias} show that by increasing the noise standard deviation the bias decreases and tends toward
the value associated to the use of the noise model of quantization (graphed using stars). Even though Gaussian noise does
not have the properties necessary to make the quantization error become a uniform random variable, it approximates such 
behavior as its variance increases. Results are also consistent with data in \cite{PetriPaglierani}, where it is shown that
for values of $\sigma/\Delta>0.3$, the overall quantization error tends to Gaussianity regardless of the ADC resolution. 
At the same time Fig.~\ref{fignoise_var} shows that the normalized maximum value of the estimation variance increases with 
the noise variance, as expected, irrespective of the quantizer resolution. 
}


{
\section{Results and Discussion}
It is high--priority of the instrumentation and measurement community to 
understand error bounds when using procedures and algorithms to analyze data.
Thus, the results in this paper serve a double role: warn against the usage of the
noise model of quantization to provide 
bounds on estimation errors when using the LS--algorithm on quantized data and 
show how to include the effect of quantization when doing calculations needed to derive such bounds.

{
\subsection{Some Application Examples}
}
Results have practical relevance.
As an example, consider 
the case when two different laboratories want to compare 
results obtained when measuring electrical parameters of the same ADC.
Research on these devices is ongoing and produces design and realizations optimizing various criteria, most generally including
resolution, speed of conversion and energy consumption. 
{ Low resolution ADCs are used 
in ultra--wide bandwidth receivers (5 bits) \cite{Ginsburg}, serial links (4.5 bits) \cite{Harwood}, hard--disk drive read channels (6 bit)
\cite{Cao} and waveform capture instruments and oscilloscopes (8 bit) \cite{Poulton}}.
{ Conversely} low--conversion rate, high resolution ADCs are employed, for instance, in { industrial instrumentation}  
or in digital audio applications. Regardless of the device resolution, all ADCs undergo testing procedures, that must be
accurate, fast and sound from a metrological point of view.  
The majority of standardized tests require sine waves as stimulating signals \cite{IEEE1241}, whose parameters
(e.g. amplitude) are { obtained} by using { LS--based estimators applied to} 
the quantized data { sequence provided by the device--under--test \cite{Pacut2}}.
Synchronizing the initial record phase of the sinusoidal signal, can be done only at the expense of added 
instrumental complexity and up to a certain uncertainty. Therefore, allowing it to freely varying 
among collected data records, as it is done frequently in practice, implies an added source of variability in the results, {
that can be modeled by assuming a random initial record phase. The sine wave amplitude is the input parameter
for the estimation of many relevant ADC parameters, such as the number of effective bits. Thus, the performed analysis
shows what to expect in the amplitude estimator properties, when the initial record phase can not be controlled over different
realizations of the same ADC testing procedure, under repeatability conditions.
The same also applies, when two laboratories verify the level of compatibility in the estimates of the sine wave amplitude,
under reproducibility conditions: the uncertainty associated to the phase variability must be taken into account.} While simulations may provide directions for 
further investigations and hints on the existence of unexpected phenomena, they must be accompanied by
analyses made to reduce the role of subjective interpretations.
Mathematically--derived bounds presented in this paper serve this scope. 
As an example, { consider the case in which} 
the normalized difference in the sine wave amplitude estimated by the two laboratories {or over different realizations
of the same ADC testing procedure} is larger than 
what can be predicted by data shown in Fig. 10, applicable as an example to a 10 bit resolution ADC.
{Then, variability in the initial record phase can not be the unique cause and
sources of uncertainty other than those associated to the effects of quantization must be looked for.}


As an additional usage example consider a medium--resolution (e.g. 10 bit) ADC embedded in a microcontroller.
The acquisition and estimation of the amplitude of a sine wave is a typical problem in many engineering areas.  
{This happens for instance} when measuring power quality{,} when performing built--in--self--test procedures to assess functional status of system--level devices { or when taking impedance measurements using sine--fitting techniques \cite{Ramos}}. Simple sinusoidal sources used frequently in this latter case, may not allow synchronized
acquisition, while synchronization with the phase of the power network in the former case, may not be feasible at reasonable costs or uncertainty levels.
In both cases, LS--based estimation of the amplitude of the sinusoid is done at signal processing level and requires assessment of the associated uncertainties. Results in Fig. 5 and 6 can be used to this aim in accordance with procedure written in \cite{GUM}, 
while corresponding expressions in the appendix can be used to cope with different 
quantizers' resolutions and records lengths.

{
\subsection{Influence of Input Signal Properties}
}
{ 
Situations occur in which tests are performed on a reduced number of output codes, that
is only few among all possible ADC codes are excited. Similarly, systems such as those in Fig.\ref{figmodel} may be sourced 
by sine waves with arbitrary large or small amplitudes within the admitted input range.
When this case, even high--resolution ADCs may induce large relative errors in amplitude estimates, as those associated to low--resolution ADCs. As an example, consider a $16$ bit ADC used in Fig.~\ref{figmodel} to quantize a 
a sine wave, whose amplitude only fully excites $256$ output codes. 
Fig.~\ref{figthree} shows that
the expected behavior of the LS--based estimator is approximately that associated to an $8$ bit ADC used at full--scale, once errors have been normalized with respect  to the width of the quantization step.


As shown in subsection~\ref{subnoise}, the addition of Gaussian noise randomizes the error 
due to the non--linear behavior of the quantizer and reduces the estimation bias. 
Zero--mean uniformly distributed noise
would have a similar or better 
linearizing behavior. With a properly set variance, 
uniform noise nulls the quantization error mean, while making the quantization
error variance, input--signal dependent ({\em noise modulation.}).
As data in Fig.~\ref{fignoise_bias} show, 
also an input offset may reduce the bias, if it randomly varies in a small--amplitude range.
In this Figure, the arithmetic mean value provides almost everywhere a much smaller estimator bias 
with respect to any of the possible values of the offset, taken as a fixed deterministic value.
The arithmetic mean is generally a reasonable estimator of the mean value of a random variable.
Thus, its behavior in Fig.~\ref{fignoise_bias}, approximates the behavior of the bias, where the input offset a random
and not a deterministic value. In this case, the offset would behave as a dither signal itself.
}

Finally, in some applications, sequential least squares are used to provide amplitude estimations over time, when  
sampling an on--going continuous--time sinusoidal waveform \cite{Kay}. Clearly, results presented here 
are applicable also in this case.

}
\vskip0.5cm
\color{black}
\section{Conclusions}
In this paper, we considered LS--based estimation of the square amplitude and {amplitude} of a quantized sine wave. 
{ The main paper contribution is summarized by results in Tab.~\ref{tabone}}.
Using several {analytical} techniques we proved that the simple noise model of quantization provides erroneous 
results under several conditions and may fail, when assessing the span of 
estimation errors. 
{ This is especially relevant when measurement results 
are used for conformance testing, to assess whether manufactured products meet specified standards.
As an example, ADC testing requires measurement of several parameters (e.g. ENOB) based on the estimation
of the properties of testing signals such as sine waves. As shown in this paper, estimates may be 
affected by relevant biases that may induce in wrong decisions 
about the device having characteristics being under or over given thresholds.} 
It has also been proved that the estimator is inconsistent, biased and that its variance is not
predicted well by the noise model of quantization. 
Exact expressions have been provided that allow a rigorous evaluation of estimator properties. 
Both the obtained results and the methods used in this paper, are applicable also when other sine wave 
parameters are estimated on the basis of quantized data { and when solving similar estimation problems.}



\numberwithin{equation}{section}

\appendices

\section{Derivation of (\ref{propsimple}) \label{appeq}}
Define
\[
R \coloneqq \frac{1}{N}\sum_{i=0}^{N-1}y_ih_i, \qquad
S \coloneqq \frac{1}{N}\sum_{i=0}^{N-1}h_i^2
\]
By using the hypothesis of zero--mean and uncorrelated random variables,
\begin{align}
&
E(R) = E\left( 
\frac{1}{N}\sum_{i=0}^{N-1}(\theta h_i^2+e_ih_i) 
\right) = 
\frac{\theta}{N}\sum_{i=0}^{N-1} E(h_i^2)=\frac{\theta}{2} \\
& 
E(S) = E\left( 
\frac{1}{N}\sum_{i=0}^{N-1}h_i^2
\right) =\frac{1}{2}
\label{a1}
\end{align}
Also
\begin{align}
E(R^2) 
& =  
\frac{1}{N^2}\sum_{i=0}^{N-1}
E(y_i^2h_i^2)+\frac{N(N-1)}{N^2}\mbox{Corr}(y_lh_l, y_kh_k)= \nonumber \\
& = 
 \frac{\theta^2}{8N}+
 \frac{\Delta^2}{24N}+
 \frac{\theta^2}{4}, \quad l \neq k
\end{align} 
and
\begin{align}
E(S^2) 
& =  
\frac{1}{N^2}\sum_{i=0}^{N-1}
E(h_i^4)+\frac{N(N-1)}{N^2}\mbox{Corr}(h^2_l, h^2_k) = \nonumber \\
& =\frac{1} {8N}+\frac{1}{4},
 \quad l \neq k.
\end{align}
Therefore
\begin{align}
\label{a4}
\begin{split}
\mbox{Var}(R)  
 = \frac{\theta^2}{8N}+
 \frac{\Delta^2}{24N},
 \qquad 
\mbox{Var}(S)  
 = \frac{1}{8N} 
\end{split}
\end{align}
and the covariance between $R$ and $S$ is given by 
\begin{align}
\mbox{Cov}(R,S) 
& =  
\frac{1}{N^2}
E\left(
\sum_{i=0}^{N-1}
(\theta h_i^2+e_ih_i)
\sum_{u=0}^{N-1}
h_u^2
\right) -\frac{\theta}{4} = \nonumber \\
& =\theta E(S^2)-\frac{\theta}{4}=\frac{\theta}{8N}
\label{ultim}
\end{align}
The expected value $E(\cdot)$ and variance $\mbox{Var}(\cdot)$ of the ratio $R/S$ of two random variables, can be approximated 
by using a Taylor series expansion as follows \cite{KendallStuart}:
\begin{align}
&
E\left(\frac{R}{S}\right)\simeq \frac{E(R)}{E(S)}\left\{
1-
\frac{
\mbox{Cov}(R,S)}{E(R)E(S)}
+\frac{\mbox{Var}(S)}{E^2(S)}
\right\}, \nonumber \\
&
\mbox{Var}\left( \frac{R}{S}\right) \simeq 
\frac{E^2(R)}{E^2(S)}\left\{
 \frac{\mbox{Var(R)}}{E^2(R)} 
-2\frac{\mbox{Cov}(R,S)}{E(R)E(S)}+
\frac{\mbox{Var}(S)}{E^2(S)}
\right\}
\label{taylor}
\end{align}
Thus, by substituting (\ref{a1})--(\ref{ultim}) in
 (\ref{taylor}), (\ref{propsimple})  follows.

\section{Amplitude Domain Approach: Moments of the Square Amplitude Estimator \label{amplitude}}
\vskip1cm

\begin{lemma}
\label{lem}
Assume $0\leq c <1$, $0 \leq \varphi \leq 1$ and $0 \leq L_1 \leq L_2 \leq 1$. Then, the solutions for
$\varphi$, of the inequality
\[
	L_1 \leq \langle c+\varphi \rangle  < L_2
\]
are
\[
\begin{array}{ll}
\{ L_1 -c \leq \varphi < L_2-c\} & c \leq L_1\\
\{0 \leq \varphi < L_2-c \} \cup \{1-c +L_1\leq \varphi \leq 1\} & L_1< c \leq L_2  \\
\{1-c+L_1 \leq \varphi < 1-c+L_2 \}& c>L_2  
\end{array}
\]
The proof is straightforward once observed that the expression  
$\langle c+\varphi \rangle$ as a function of $\varphi$,  is piecewise linear:
\[
\langle c+\varphi \rangle =
\left\{
\begin{array}{ll}
\varphi+c & 0 \leq \varphi < 1-c \\
\varphi-(1-c) & 1-c \leq \varphi < 1 
\end{array}
\right.
\]
\end{lemma}
\subsection{A Model for the Quantizer Output}
 \label{l1}
{ 
The model described in this Appendix is based on a similar approach taken in \cite{CarbonePetri}.
However, it extends it in several ways: 
it presents a more strict mathematical formalization 
of the quantizer output, it includes a mathematical description of the quantizer 
higher--order statistical moments and it proves its applicability
to the analysis of mean value and correlation 
of quantized stochastic processes. 
}
Assume 
\[
	\overline{x}_u(\varphi)\coloneqq -\frac{A}{\Delta}\cos(k_u+\varphi)+0.5 {+ c}, \qquad u=0, \ldots, N-1
\]
with $A$, $\Delta$ { and $c$} real numbers, $k_u$ a sequence of real numbers,
$N$ a positive integer and $0\leq \varphi \leq 2\pi$. 
{The constant $c$ models both the contribution of an offset in 
the sinusoidal signal and of
an eventual offset in the quantizer input--output characteristic, as they are indistinguishable.
In fact, while in this paper, mid--tread quantization is considered, by setting properly the value of
$c$, other types of quantization are covered by this analysis. For instance, 
when $c=-0.5$, a truncation quantizer is modeled. Conversely by setting $c=\frac{d}{\Delta}$,
the following analysis also comprehends the case 
when the input sine wave has an offset $d$.}
Define the quantizer output as 
\[
	y_u\coloneqq \Delta \lfloor \overline{x}_u \rfloor, \qquad u=0, \ldots, N-1.
\]
Then, 
\[
	y_u =n\Delta,  \qquad n \in (-\infty, \infty), \qquad n \quad \mbox{integer}{,}
\]
if $\overline{x}_u(\varphi)$ belongs to the interval:
\[
S_n\coloneqq \left[D_n, U_n \right),
\]
where $D_n\coloneqq \max(\min_{\varphi}\{\overline{x}_u(\varphi)\},  n)$ and
$U_n\coloneqq \min(\max_{\varphi}\{\overline{x}_u(\varphi)\}, n+1)$,
Consequently,
\[
y_u=\sum_{n =-\infty}^{\infty} n\Delta i\left( \overline{x}_u(\varphi) \in S_n \right), \qquad S_n \coloneqq  [D_n, U_n)
\]
where $i(\cdot)$ is the indicator function of the event at its argument.
For a given value of $k_u$, $\overline{x}_u(\varphi)$ will or will not 
belong to $S_n$ depending on the value of $\varphi$. 
In the following, it will be shown how to find the set of values that make
$y_u=n\Delta$. 
This occurs when
\be
 D_n \leq -\frac{A}{\Delta}\cos \left( k_u+\varphi \right) +0.5 {+c}<  U_n
\label{se}
\ee
that is
\[
	\frac{\Delta}{A}\left(D_n -0.5{-c}\right)
	\leq 
	-\cos\left( k_u+\varphi \right) 
       < 
	\frac{\Delta}{A}\left(  U_n -0.5{-c}\right)
\]
Since the cosine function is periodic with period $2\pi$, we may write:
\begin{align*}
	\frac{\Delta}{A}\left( D_n-0.5{-c}\right)
	& \leq 
	-\cos  \left( \left \langle \frac{k_i+\varphi}{2\pi}\right \rangle 2\pi \right) < \\
	&
       < 
	\frac{\Delta}{A}\left(   U_n  -0.5{-c}\right)
\end{align*}
Thus,
\begin{align}
\begin{split}
-\frac{\Delta}{A}\left( U_n  -0.5{-c}\right)
		 & < 
	\cos\left( \left \langle \frac{k_u+\varphi}{2\pi}\right \rangle 2\pi \right) \leq \\
&       \leq 
-\frac{\Delta}{A}\left( D_n -0.5{-c}\right)
\label{ddd}
\end{split}
\end{align}
Define 
\[
L_n \coloneqq -\frac{\Delta}{A}\left(D_n -0.5{-c}\right),
\qquad
R_n\coloneqq -\frac{\Delta}{A}\left( U_n -0.5{-c}\right),
\]
and observe that the $\arccos(\cdot)$ function is a decreasing function of its argument and returns values in $[0,\pi]$, so that by applying it to all members in (\ref{ddd}) we obtain:
%
\begin{align*}
& \left\{
\frac{1}{2\pi}\arccos(L_n) \leq \left \langle \frac{k_u+\varphi}{2\pi}\right \rangle < \frac{1}{2\pi}\arccos(R_n)
\right\} \cup \\
& \cup
\left\{
\left( 1-\frac{1}{2\pi}\arccos(R_n)\right)
< \left \langle \frac{k_u+\varphi}{2\pi}\right \rangle
\leq \right. \\
& 
\left. 
\quad \leq
\left( 1-\frac{1}{2\pi}\arccos(L_n)\right)
\right\}
\end{align*}
Moreover, since $\left \langle a+b \right \rangle = \left \langle \langle a \rangle+ \langle b \rangle \right \rangle$ and $0 \leq \varphi < 2\pi$, for $u=0, \ldots, N-1$, we have
\begin{align}
& \left\{
\frac{1}{2\pi}\arccos(L_n) \leq \left \langle 
\left \langle \frac{k_u}{2\pi}\right \rangle+
\frac{\varphi}{2\pi}\right \rangle < \frac{1}{2\pi}\arccos(R_n)
\right\}
\cup \nonumber \\
& \cup 
\left\{
1-\frac{1}{2\pi}\arccos(R_n)
< \left \langle \left \langle \frac{k_u}{2\pi}\right \rangle+\frac{\varphi}{2\pi}\right \rangle \leq
\right. \nonumber \\
& \left. \quad \leq
1-\frac{1}{2\pi}\arccos(L_n)
\right\}
\label{fe}
\end{align}
Let us define $\Phi_n(k_u)$, for a given value of $k_u$, 
as the sets of values of $\varphi$ such that (\ref{ddd}) is satisfied.
This set is to be found by applying twice
the results in lemma (\ref{lem}) to 
the two intervals $I_1, I_2$, implicitly defined in (\ref{fe})
\begin{align*}
&
I_1 \coloneqq \left[
\frac{1}{2\pi}\arccos(L_n),
 \frac{1}{2\pi}\arccos(R_n)
\right) \\
&
I_2 \coloneqq \left(
1- \frac{1}{2\pi}\arccos(R_n),
 1- \frac{1}{2\pi}\arccos(L_n)
\right]
\end{align*}
with $c=\left \langle \frac{k_u}{2\pi}\right \rangle$, $u=0, \ldots, N-1$.
Thus,
\[
y_u=\sum_{n =-\infty}^{\infty} n\Delta i\left( \varphi \in \Phi_n(k_u) \right)
\]
The set of $\Phi_n(k_u)$, $n=-\infty \ldots \infty$,
for any given $k_u$, forms a partition of the interval $[0, 2\pi)$.
\begin{figure}[t]
\begin{center}
\includegraphics[scale=0.4]{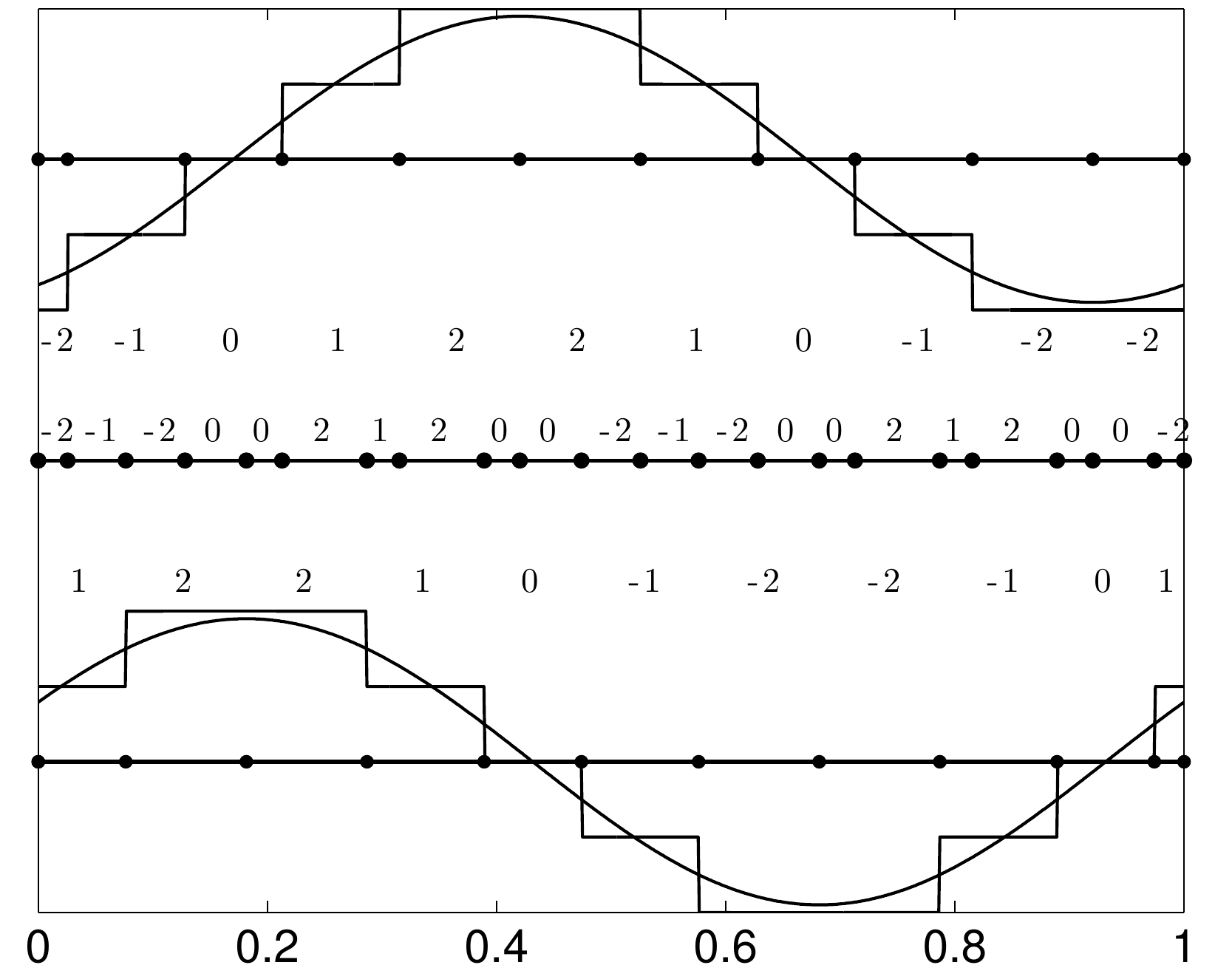}
\label{corrfig}
\caption{Usage example of the amplitude--domain model to calculate the correlation between two quantizer outputs ($b=3$), when 
$\varphi/(2\pi)$ is uniform in $[0, 1)$ { and $A=1$ and $c=0$}: the normalized products of the two outputs (center line), must be multiplied with the width
of the corresponding segment on the center line, that represents the probability of occurrence, until all contributions are summed.}
\end{center}
\end{figure}

Consider the product $y_{u_1} y_{u_2} \cdots y_{u_M}$, with integers
$u_{\cdot}$ all taking values in $\{0\ldots N-1\}$.
Then 
\begin{align*}
	&
	y_{u_1}y_{u_2} \cdots y_{u_M}  =
	\sum_{n =-\infty}^{\infty} n\Delta i\left( \overline{x}_{u_1}(\varphi) \in S_n \right) \cdot \\ 
	&
	\cdot
	\sum_{m =-\infty}^{\infty} m\Delta i\left( \overline{x}_{u_2}(\varphi) \in S_m \right) \cdots 
	\sum_{h =-\infty}^{\infty} h\Delta i\left( \overline{x}_{u_M}(\varphi) \in S_h \right). 
\end{align*}
Since the intervals $S_n$ and $S_m$ have void intersection, if $m \neq n$, we can write
\begin{multline*}
	y_{u_1}y_{u_2} \cdots y_{u_M} =\Delta^M
	\sum_{n =-\infty}^{\infty} n^M i\left(\overline{x}_{u_1}(\varphi) \in S_n \right) \\
\cdot
 i\left( \overline{x}_{u_2}(\varphi) \in S_n \right) \cdots
 i\left( \overline{x}_{u_M}(\varphi) \in S_n \right)
\end{multline*}
that is,
\begin{multline*}
y_{u_1}y_{u_2} \cdots y_{u_M} =\Delta^M
\sum_{n =-\infty}^{\infty} n^M i\left(\varphi \in \Phi_n(k_{u_1})\right) \\
\cdot
 i\left( \varphi \in \Phi_n(k_{u_2})\right) \cdots
 i\left(\varphi \in \Phi_n(k_{u_M}) \right)
\end{multline*}
Given that the indicator function can only output $0$ or $1$,
by the same reasoning, we have:
\begin{multline}
y^{m1}_{u_1}y^{m2}_{u_2} \cdots y^{m_M}_{u_M} =\Delta^{m_1+m_2+\cdots +m_M}  \\
\cdot \sum_{n =-\infty}^{\infty} n^{m_1+m_2+\cdots +m_M} 
	 i\left(\varphi \in \Phi_n(k_{u_1})\right) \\ \cdot
 i\left( \varphi \in \Phi_n(k_{u_2})\right) \cdots
 i\left(\varphi \in \Phi_n(k_{u_M}) \right)
\label{moment}
\end{multline}
where $m_1, m_2, \cdots, m_M$ are integer values.
Thus, the product
 \[
 	y^{m_1}_{u_1}y^{m2}_{u_2} \cdots y^{m_M}_{u_M}
\] 
is a deterministic function of 
$\varphi$. 
If $\varphi$ becomes a random variable distributed in $[0, 2\pi)$, then 
(\ref{moment}) allows the calculation of joint moments of the quantizer output as follows
\begin{multline*}
E\left(
y^{m_1}_{u_1}y^{m2}_{u_2} \cdots y^{m_M}_{u_M}
\right) =
\Delta^{m_1+m_2+\cdots +m_M} \\
\cdot
	\sum_{n =-\infty}^{\infty} n^{m_1+m_2+\cdots +m_M}\cdot E\left(
i\left(\varphi \in \Phi_n(k_{u_1})\right) \right.  \\
\cdot
\left.
 i\left( \varphi \in \Phi_n(k_{u_2})\right) \cdots
 i\left(\varphi \in \Phi_n(k_{u_M}) \right)\right)
\end{multline*}
The argument of the expectation in the summation is a Bernoulli random variable. 
Therefore,
\begin{multline}
E\left(
y^{m_1}_{u_1}y^{m2}_{u_2} \cdots y^{m_M}_{u_M}
\right) =
\Delta^{m_1+m_2+\cdots +m_M}  \\
\cdot
	\sum_{n =-\infty}^{\infty} n^{m_1+m_2+\cdots +m_M} \cdot \mbox{Pr}\left(
i\left(\varphi \in \Phi_n(k_{u_1})\right) \right. \\
\left. \cdot
 i\left( \varphi \in \Phi_n(k_{u_2})\right) \cdots
 i\left(\varphi \in \Phi_n(k_{u_M}) \right) = 1\right)
\label{cc}
\end{multline}
Define
\[
\Phi \coloneqq \Phi_n(k_{u_1}) \cap \Phi_n(k_{u_M}) \cap \cdots \cap \Phi_n(k_{u_M}),
\]
that may be the union of disjoint intervals.
Then, if $\varphi$ is uniformly distributed, 
the probability in (\ref{cc})
can be calculated as the $L_1$ norm of $\Phi$ normalized to $2\pi$, that is $|| \Phi ||_1/(2\pi)$.

To illustrate how the model can be used to calculate the correlation 
between two quantizer outputs, 
consider the sine waves depicted in Fig. 11
as a function
of $0\leq \frac{\varphi}{2\pi} < 1$, assuming two different values of $k_{u_1}$ and $k_{u_2}$.
The $2$ sine waves are quantized using a 3 bit quantizer: below and above 
the upper and lower sine waves, respectively, is printed the corresponding quantized value, normalized to $\Delta$.
The product of these two sequences is written on the center line, in this Figure.  
Since $\varphi$ is uniformly distributed in $[0, 2\pi)$, the correlation can be 
obtained by multiplying  the magnitude of each segment in the center line, 
by the corresponding integer value printed above it.


\section{Frequency Domain Approach: Bias of the Square Amplitude Estimator \label{appbias}}
{The model described in this Appendix is based on a Fourier--series expansion of the
quantization error sequence, a technique also used elsewhere to find mathematical 
expressions and properties of quantized sequences \cite{KollarBook}. It is here applied, together 
with original research results, to provide a fully developed example of its applicability to solve
LS--based estimation problems, based on quantized data.
}
\begin{theorem}
Consider the product
\begin{multline}
	C(I,U,H,L,\varphi)  \coloneqq
	\cos(I(k_i+\varphi))	
	\cos(U(k_u+\varphi)) \\
	\cdot
	\cos(H(k_h+\varphi))
	\cos(L(k_l+\varphi))
	\label{cphi}
\end{multline}
where $k_n\coloneqq \frac{2\pi}{N}\lambda n$, $n = 0,\ldots, N-1$,
$\{I, U, H, L\}$ is a set of positive odd integer numbers and $\varphi$ is a random variable uniform in $(0,2\pi]$. 
By using trigonometric identities 
(\ref{cphi}) becomes:
\begin{align}
\label{ridda}
	&
	C(I,U,H,L,\varphi)= \nonumber \\ 
	&
	=\frac{1}{8}\left\{
	\cos\left(Ik_i+Uk_u+Hk_h+Lk_l+(I+U+H+L)\varphi \right)\right. \nonumber \\
	&
	\quad + \cos\left(Ik_i+Uk_u-Hk_h-Lk_l+(I+U-H-L)\varphi \right)  \nonumber \\
	&
	\quad + \cos\left(Ik_i+Uk_u+Hk_h-Lk_l+(I+U+H-L)\varphi \right)\nonumber \\
	&
	\quad +\cos\left(Ik_i+Uk_u-Hk_h+Lk_l+(I+U-H+L)\varphi \right) \nonumber \\
	&
	\quad +\cos\left(Ik_i-Uk_u+Hk_h+Lk_l+(I-U+H+L)\varphi \right) \nonumber \\
	& 
	\quad +\cos\left(Ik_i-Uk_u-Hk_h-Lk_l+(I-U-H-L)\varphi \right) \nonumber \\ 
	&
	 \quad +\cos\left(Ik_i-Uk_u+Hk_h-Lk_l+(I-U+H-L)\varphi \right)\nonumber \\
	&
	\quad
	\left.
	+ \cos\left(Ik_i-Uk_u-Hk_h+Lk_l+(I-U-H+L)\varphi \right) 
	\right\}
\end{align}
Since $C(\cdot, \cdot, \cdot, \cdot, \varphi)$ is periodic in $\varphi$ with period $2\pi$,
its expected value is different
from $0$, only when at least one of  the coefficients of $\varphi$ is $0$. 
This produces a set of Diophantine equations in the form:
\be
I \pm U = \pm H \pm L 
\label{ULHK}
\ee
Given that (\ref{ULHK}) is satisfied, and assuming $I$, $U$, $H$ and $L$ to be positive integers,
the sums
\be
\sum_{i=0}^{N-1}
\sum_{u=0}^{N-1}
\sum_{h=0}^{N-1}
\sum_{l=0}^{N-1}
E(C(I,U,H,L,\varphi))\cos(k_i-k_u)\cos(k_h-k_l)
\label{origin}
\ee
do not vanish only if $I=U=H=L=1$.
\begin{proof}
Using (\ref{ridda}) and (\ref{ULHK}),  (\ref{origin}) becomes the summation 
of terms of the type
\begin{multline}
c(k_i,k_u,k_h,k_l)  \coloneqq \mbox{coefficient}  \\
\cdot \cos(Ak_i+Bk_u+Ck_h+Dk_l)\cos(k_i-k_l)\cos(k_h-k_l),
\label{coe}
\end{multline}
where $A,B,C,D$ are positive or negative odd integers.
Any of the expressions of the type in (\ref{coe}) is summed over the four indices $i,u,h,l$.
Consider, for instance the summation over $i$. By using trigonometric identities we have:
\begin{multline}
\label{ll0}
\sum_{i=0}^{N-1}c(k_i,k_u,k_h,k_l)  = 
\mbox{coefficient}  \\
\cdot \sum_{i=0}^{N-1}
\left\{ 
\cos(Ak_i)\cos(Bk_u+Ck_h+Dk_l) 
\right.\\
\cdot
\cos(k_i)\cos(k_l)\cos(k_h-k_l) \\
  +\cos(Ak_i)\cos(Bk_u+Ck_h+Dk_l) \\
\cdot \sin(k_i)\sin(k_l)\cos(k_h-k_l) \\
 -\sin(Ak_i)\sin(Bk_u+Ck_h+Dk_l)\\
\cdot \cos(k_i)\cos(k_l)\cos(k_h-k_l) \\
-\sin(Ak_i)\sin(Bk_u+Ck_h+Dk_l)  \\
\left.
\cdot
\sin(k_i)\sin(k_l)\cos(k_h-k_l)
\right\}.
\end{multline}
Consider the first term in the summation. By the Euler's formula we have:
\begin{multline}
\sum_{i=0}^{N-1}
 \cos(Ak_i)\cos(Bk_u+Ck_h+Dk_l) \\
\cdot \cos(k_i)\cos(k_l)\cos(k_h-k_l) \\
=c_1\sum_{i=0}^{N-1}
\left(
\frac{e^{jA\frac{2\pi}{N}\lambda i}-e^{-jA\frac{2\pi}{N}\lambda i}}{2}
\frac{e^{j\frac{2\pi}{N}\lambda i}-e^{-j\frac{2\pi}{N}\lambda i}}{2}\right)\\
=\frac{c_1}{4}\left(
\frac{1-e^{j(A+1)2\pi \lambda}}{1-e^{j(A+1)\frac{2\pi}{N} \lambda}}
+
\frac{1-e^{j(A-1)2\pi \lambda}}{1-e^{j(A-1)\frac{2\pi}{N} \lambda}} +\right.\\
\left.
+
\frac{1-e^{-j(A+1)2\pi \lambda}}{1-e^{-j(A+1)\frac{2\pi}{N} \lambda}}
+
\frac{1-e^{-j(A-1)2\pi \lambda}}{1-e^{-j(A-1)\frac{2\pi}{N} \lambda}}
\right),
\label{ll}
\end{multline}
where $c_1$ is constant with respect to $i$ and the latter equality
follows by the geometric sum formula.
Since $\lambda$ and $A$ are integers,
all terms in (\ref{ll}) vanish unless $A = \pm 1$.  
In both cases, (\ref{ll}) results in $c_1\frac{N}{2}$.
The same reasoning applies to the $4-th$ term in (\ref{ll0}), that is the
product of all sine wave functions, while the cross--product terms
of the type: $\mbox{constant}\cdot\sin(Ak_i)\cos(k_i)$ and $\mbox{constant}\cdot\cos(Ak_i)\sin(k_i)$, 
when summed over $i$, vanish regardless of $A$.
Since this argument applies for any of the indices in (\ref{origin}) the lemma is proved.
Then, under the assumption $N>2$,  the sums
\begin{multline}
\sum_{i=0}^{N-1}
\sum_{u=0}^{N-1}
\sum_{h=0}^{N-1}
\sum_{l=0}^{N-1}
E\left(
C(1,1,1,1,\varphi)
\right) \\
\cdot 
\cos(k_i-k_u)\cos(k_h-k_l)=\frac{N^4}{16},
\label{se}
\end{multline}
where the last equality follows by
 expanding each term using the Euler's formula and by summing the corresponding geometric sums.
\end{proof}
\end{theorem}
Thus, from (\ref{estima}) we can write:
\begin{multline*}
\hat{A}^2  = \frac{4}{N^2}
\sum_{i=0}^{N-1}\sum_{u=0}^{N-1} (s_i+e_i(s_i))(s_u+e_u(s_u)) \cos\left( k_i-k_u\right) \\ 
=
\frac{4}{N^2}\sum_{i=0}^{N-1}\sum_{u=0}^{N-1}z_{iu} \cos\left( k_i-k_u \right)
\end{multline*}
where
\begin{align}
\begin{split}
z_{iu} \coloneqq & s_is_u +e_i(s_i)s_u +e_u(s_u)s_i +e_i(s_i)e_u(s_u)  \\
& 
\quad \qquad
i=0, \ldots, N-1 \quad u=0, \ldots, N-1 
\label{yij}
\end{split}
\end{align}
Each term in (\ref{yij}) is a deterministic function of the random variable $\varphi$ and
will be analyzed separately in the following.
With $\varphi$ uniform  in $[0, 2\pi)$, we have
\begin{align*}
	E(s_is_u) = \frac{A^2}{2} & \cos(k_i-k_u)\\
	& i=0, \ldots, N-1,  \quad u=0,\ldots, N-1
\end{align*}
Since, when $N>2$
\[
\frac{4}{N^2}\sum_{i=0}^{N-1}\sum_{u=0}^{N-1} \cos^2(k_i-k_u) = 2,
\]
then,
\[
\frac{4}{N^2}\sum_{i=0}^{N-1}\sum_{u=0}^{N-1}E(s_is_u) \cos(k_i-k_u) = A^2.
\]
Consider now the term $e_i(s_i)s_u$. We have \cite{KollarBook}
\begin{align}
\begin{split}
e_i(s_i)& = \sum_{k=1}^{\infty}(-1)^k\frac{\Delta}{\pi k}\sin \left( 2\pi \frac{k}{\Delta} s_i\right) 
\\
&
=\sum_{k=1}^{\infty}(-1)^k\frac{\Delta}{\pi k}\sin \left( -2A\pi \frac{k}{\Delta} \cos\left( k_i+\varphi \right) \right) \\
& = \sum_{k=1}^{\infty}(-1)^k\frac{2\Delta}{\pi k} \sum_{h=0}^{\infty}(-1)^hJ_{2h+1}\left( -\frac{2\pi k A}{\Delta}\right) \\
&
\hskip3.2cm \cdot \cos\left( (2h+1)(k_i+\varphi) \right)
\label{error}
\end{split}
\end{align}
where the last equality follows by observing that \cite{AbramovitzStegun}
\be
\sin\left( z \cos(\beta)\right) =2 \sum_{h=0}^{\infty}(-1)^hJ_{2h+1}(z)\cos\left((2h+1)\beta\right)
\label{sinj}
\ee
Thus
by defining
$
z_h \coloneqq \frac{2\pi h A}{\Delta}$, $h=0, 1, \ldots$
we have
\begin{multline}
\frac{4}{N^2}\sum_{i=0}^{N-1}\sum_{u=0}^{N-1}E(e_i(s_i)s_u) \cos(k_i-k_u)  \\
=-\frac{4A}{N^2}
\sum_{i=0}^{N-1}
\sum_{u=0}^{N-1}E\left( 
\sum_{k=1}^{\infty}(-1)^k\frac{2\Delta}{\pi k}
\sum_{h=0}^{\infty}(-1)^hJ_{2h+1}
\left( -z_k\right)\right. \\
\cdot
\left.
\cos\left( (2h+1)(k_i+\varphi) \right)
\cos(k_i+\varphi)
\cos(k_i-k_u) 
\right) \\
=-\frac{4A}{N^2}
\sum_{i=0}^{N-1}
\sum_{u=0}^{N-1}
\sum_{k=1}^{\infty}(-1)^k
\frac{2\Delta}{\pi k} 
\sum_{h=0}^{\infty}(-1)^hJ_{2h+1}
\left( -z_k\right) \cdot \\
\cdot
E
\left( 
C(2h+1,1,0,0,\varphi) 
\right)
\cos(k_i-k_u)
\label{es}
\end{multline}
where 
the last equality holds by virtue of the dominated convergence theorem \cite{KolmogorovFomin}.
The expected value in (\ref{es}) does not vanish only if $h=0$, when we have:
\[
	E\left( 
C(1,1,0,0,\varphi) 
\right) = \frac{1}{2}\cos(k_i-k_u)
\]
Thus, from (\ref{error}), we have: 
\be
E(e_i(s_i)s_u)= 
\frac{A\Delta}{\pi}
\cos(k_i-k_u)\sum_{k=1}^{\infty} \frac{(-1)^k}{k}J_1\left( z_k\right)
\label{cross}
\ee
that is equal to $E(e_u(s_u)s_i)$ because of the cosine function being an even function of its argument and
$J_1(\cdot)$ being an odd function of its argument.
By the same reasoning, we have: 
\[
	\sum_{i=0}^{N-1}\sum_{u=0}^{N-1} 
	E
\left( 
C(1,1,0,0,\varphi) 
\right)
\cos(k_i-k_u)= \frac{N^2}{4}
\]
Consequently, (\ref{es}) becomes:
\begin{multline}
\frac{4}{N^2}\sum_{i=0}^{N-1}\sum_{u=0}^{N-1}E(e_i(s_i)s_u)  \cos(k_i-k_u)  \\
=\frac{2A\Delta}{\pi}
\sum_{k=1}^{\infty} \frac{(-1)^k}{k}J_1\left( z_k\right)
\label{es2}
\end{multline}
Now consider the term $e_i(s_i)e_u(s_u)$ in (\ref{yij}). We have:
\begin{multline}
e_i(s_i)e_u(s_u)  = \left( \frac{\Delta}{\pi} \right)^2 
\sum_{k=1}^{\infty}
\sum_{h=1}^{\infty}   
\frac{(-1)^{h+k}}{h k}  \\
\cdot
\sin \left(2 \pi \frac{k}{\Delta} s_i\right)
\sin \left(2 \pi \frac{h}{\Delta} s_u \right) 
\label{ee}
\end{multline}
By using (\ref{sinj}), the rightmost product in (\ref{ee}) 
has the following expected value:
\begin{multline}
E
\left(
\sin \left(2 \pi \frac{k}{\Delta} s_i\right)
\sin \left(2 \pi \frac{h}{\Delta} s_u \right)
\right) 
 \\
=4  \sum_{m=0}^{\infty} \sum_{n=0}^{\infty}(-1)^{m+n} J_{2m+1}(z_k)
J_{2n+1}(z_h)\\ 
 \cdot
E
\left(
C(2m+1,2n+1,0,0, \varphi)
\right)
\label{qwer}
\end{multline}
By neglecting negative values of the indices, because of (\ref{ULHK}), 
the expectation in (\ref{qwer}) is different from zero only if $m=n$, when we have 
\begin{multline}
E \left(
\sin \left(2 \pi \frac{k}{\Delta} s_i
\right)
\sin \left( 2 \pi \frac{h}{\Delta} s_u 
\right)
\right) \\
= 2\sum_{n=0}^{\infty} J_{2n+1}(z_k)
J_{2n+1}(z_h) \cos((2n+1)(k_i-k_u))
\label{eee}
\end{multline}
which corresponds to the analysis done in \cite{KokkelerGunst}\cite{Hurd} and to the results published in 
App.~G, in \cite{KollarBook}.
By using (\ref{ee}) and (\ref{eee}) we obtain:
\begin{multline}
\frac{4}{N^2}\sum_{i=0}^{N-1}\sum_{u=0}^{N-1}
E\left( 
e_i(s_i)e_u(s_u)
\right)\cos(k_i-k_u) \\
=\frac{8\Delta^2}{\pi^2 N^2}\sum_{i=0}^{N-1}\sum_{u=0}^{N-1}
\cos(k_i-k_u)
\sum_{k=1}^{\infty}
\sum_{h=1}^{\infty}   
\frac{(-1)^{h+k}}{h k}   \\
\cdot
\sum_{n=0}^{\infty}
J_{2n+1}(z_h)
J_{2n+1}(z_k)
\cos((2n+1)(k_i-k_u)).
\label{newbb}
\end{multline}
By using (\ref{newbb}) and twice (\ref{es2}), one can write:
\begin{multline}
E({\hat A^2}) = 
A^2+\frac{4A\Delta}{\pi}
\sum_{k=1}^{\infty} \frac{(-1)^k}{k}J_1\left( z_k\right) \\
+\frac{8\Delta^2}{\pi^2 N^2}\sum_{i=0}^{N-1}\sum_{u=0}^{N-1}
\cos(k_i-k_u)
\sum_{k=1}^{\infty}
\sum_{h=1}^{\infty}   
\frac{(-1)^{h+k}}{h k}  \\
\cdot
\sum_{n=0}^{\infty}
J_{2n+1}(z_h)
J_{2n+1}(z_k)
\cos((2n+1)(k_i-k_u))
\label{ff}
\end{multline}
Expression (\ref{ff}) shows that the bias of the estimator of the square amplitude comprises two terms
as follows:
\[
bias(A,\Delta, N) \coloneqq E({\hat A}^2)-A^2=4Ag(A,\Delta)+8h(A,\Delta,N)
\]
where 
\be
g(A,\Delta) \coloneqq \frac{\Delta}{\pi}
\sum_{k=1}^{\infty} \frac{(-1)^k}{k}J_1\left( z_k\right)
\label{gg}
\ee
and
\begin{multline}
h(A,\Delta,N) \coloneqq
\frac{\Delta^2}{\pi^2 N^2}\sum_{i=0}^{N-1}\sum_{u=0}^{N-1}
\cos(k_i-k_u)  \\
 \cdot
\sum_{k=1}^{\infty}
\sum_{h=1}^{\infty}   
\frac{(-1)^{h+k}}{h k}  
\sum_{n=0}^{\infty}
J_{2n+1}(z_h)
J_{2n+1}(z_k) \\
\cdot 
\cos((2n+1)(k_i-k_u))
\label{hh}
\end{multline}
Observe that, while $g(\cdot, \cdot)$ does not depend on the number of samples, $h(\cdot, \cdot, \cdot)$
is also a function of $N$. Thus, the estimator of the square amplitude, based on least squares of quantized data is
not consistent, since its bias does not vanish when $N \rightarrow \infty$, for a finite quantizer resolution. 

Observe also that
(8.531 in \cite{TablesIntegrals})
\[
J_0(mR)=J_0(m\rho)J_0(mr)+2\sum_{k=1}^{\infty}J_k(m\rho)J_k(mr)\cos(k\phi),
\]
where,
\[
R=\sqrt{r^2+\rho^2-2r\rho\cos(\phi)}
\]
As a consequence, the rightmost summation in (\ref{hh}) can be written as follows:
\begin{multline}
\sum_{n=0}^{\infty} J_{2n+1}(z_k)
J_{2n+1}(z_h)
\cos((2n+1)
(k_i-k_u)) \\
=\frac{1}{4}(J_0(R)-J_0(\overline{R}))
\end{multline}
with
\[
R=\sqrt{z_h^2+z_k^2-2z_hz_k\cos(k_i-k_u)}
\]
and
\[
\overline{R}=\sqrt{z_h^2+z_k^2+2z_hz_k\cos(k_i-k_u)}
\]

To further characterize the bias, it is shown next the asymptotic
behavior of $h(\cdot, \cdot, \cdot)$: 
\begin{multline}
\lim_{N \rightarrow \infty}
h(A,\Delta,N)=
\lim_{N \rightarrow \infty}
\frac{\Delta^2}{\pi^2 N^2}
\sum_{i=0}^{N-1}\sum_{u=0}^{N-1}
\cos(k_i-k_u) \\
 \cdot
\sum_{k=1}^{\infty}
\sum_{h=1}^{\infty}   
\frac{(-1)^{h+k}}{h k}  
\sum_{n=0}^{\infty}
J_{2n+1}(z_h)
J_{2n+1}(z_k) \\
\cdot
\cos((2n+1)(k_i-k_u)) \\
=
\lim_{N \rightarrow \infty}
\frac{\Delta^2}{\pi^2}
\sum_{k=1}^{\infty}
\sum_{h=1}^{\infty}   
\frac{(-1)^{h+k}}{h k}  
\sum_{n=0}^{\infty}
J_{2n+1}(z_h)
J_{2n+1}(z_k)  \\
 \cdot
\left[
\frac{1}{N^2}
\sum_{i=0}^{N-1}\sum_{u=0}^{N-1}
\cos((2n+1)(k_i-k_u))
\cos(k_i-k_u)\right]
\label{nnewbb}
\end{multline}
By using Euler's formula it can be proven
that when $N \ge 3$, 
the double rightmost summation in square brackets in (\ref{nnewbb}),
is $0$, unless $n$ equals $0$,
when it becomes equal to $1/2$.
Thus (\ref{nnewbb}) becomes
\begin{multline}
\lim_{N \rightarrow \infty}
h(A,\Delta,N)
 =
\frac{1}{2}
\frac{\Delta^2}{\pi^2}
\sum_{k=1}^{\infty}
\sum_{h=1}^{\infty}   
\frac{(-1)^{h+k}}{h k}  
J_{1}(z_h)
J_{1}(z_k)
\\
=
\frac{1}{2}
\left(
\frac{\Delta}{\pi}
\sum_{h=1}^{\infty}   
\frac{(-1)^{h}}{h}  
J_{1}(z_h)
\right)^2
=  \frac{g^2(A,\Delta)}{2}
\end{multline}
Consequently, when $N\rightarrow \infty$
\be
bias(A,\Delta,N) \rightarrow bias(A,\Delta) \coloneqq 4g(A,\Delta) \left[A+g(A,\Delta)\right].
\label{bias}
\ee

\section{Frequency Domain Approach: Variance of the Square Amplitude Estimator \label{appvar}}
In this appendix, we provide expressions for the variance of the square amplitude estimator and
we analyze its asymptotic behavior when $N\rightarrow \infty$.
From (\ref{estima}), we have:
\begin{multline}
\left( \hat{A}^2\right)^2=
\frac{16}{N^4}
\sum_{i=0}^{N-1}
\sum_{u=0}^{N-1} 
\sum_{h=0}^{N-1}
\sum_{l=0}^{N-1}
z_{iuhl}
\cos\left( k_i-k_u \right) \\
\cdot 
\cos\left( k_h-k_l \right)
\label{ssum}
\end{multline}
where 
$
 z_{iuhl}\coloneqq y_iy_uy_hy_l.
$
By expanding each output 
sequence as the sum of the input and quantization error
we obtain
\begin{align*}
z_{iuhl}
& = (s_i+e_i(s_i))
(s_u+e_u(s_u))
(s_h+e_h(s_h))
(s_l+e_l(s_l))  \\
& =
z_4+z_3+z_2+z_1+z_0,
\end{align*}
where
\begin{align}
\begin{split}
z_4 \coloneqq & s_is_us_hs_l \\
z_3 \coloneqq & s_is_us_he_l(s_l)+s_is_ue_h(s_h)s_l+s_ie_u(s_u)s_hs_l \\
& +e_i(s_i)s_us_hs_l \\
z_2\coloneqq & s_is_ue_h(s_h)e_l(s_l)+s_ie_u(s_u)s_he_l(s_l)\\
&
+ s_ie_u(s_u)e_h(s_h)s_l+e_i(s_i)s_us_he_l(s_l)\\
& + e_i(s_i)s_ue_h(s_h)s_l+e_i(s_i)e_u(s_u)s_hs_l \\
z_1 \coloneqq & s_ie_u(s_u)e_h(s_h)e_l(s_l)+e_i(s_i)s_ue_h(s_h)e_l(s_l)\\
&
+ e_i(s_i)e_u(s_u)s_he_l(s_l)+e_i(s_i)e_u(s_u)e_h(s_h)s_l \\
z_0 \coloneqq & e_i(s_i)e_u(s_u)e_h(s_h)e_l(s_l) 
\end{split}
\label{yys}
\end{align}
Any of the terms in (\ref{yys}) depend on $N$, $\varphi$ and $\lambda$.
The expected value of (\ref{ssum}) is the sum of all expected values of the terms $z_{i}$, $i=0, 1, 2, 3, 4$, which in turn 
is the sum of all expected values included in their definitions above. 
The term $z_4$, when multiplied by $\frac{16}{N^4}\cos(k_i-k_u)\cos(k_h-k_l)$ and summed over the four indices $i,u,h,l$ 
will
provide the value $A^4$. 
Each term in (\ref{yys}) is the product of $4$ factors that are either the input signal or the quantization error.
Their analysis  can be then carried out in a similar way independently on the number of times the signal or the quantization error appears in each term. As an example, it is shown how to find an expression for the term
$s_ie_u(s_u)s_he_l(s_l)$ in $z_2$, in (\ref{yys}).
Because of (\ref{error}) and (\ref{sinj}) we have
\begin{multline}
s_i 		
=-A\cos(k_i+\varphi),  \quad  k_i \coloneqq  \lambda\frac{2\pi i}{N}\\
e_u(s_u) 	 = 2\frac{\Delta}{\pi} 
\sum_{k=1}^{\infty}
\sum_{r=0}^{\infty}
\frac{(-1)^{k+r}}{ k}
J_{2r+1}
\left( z_k\right) \\
\cdot 
\cos\left((2r+1)(k_u+\varphi)\right), \quad  k_u \coloneqq \lambda\frac{2\pi u}{N},   z_k \coloneqq 2\pi k\frac{A}{\Delta}  \\
s_h 		 =-A\cos(k_h+\varphi)   \quad k_h \coloneqq \lambda\frac{2\pi h}{N} \\
e_l(s_l) 	 = 2\frac{\Delta}{\pi} 
\sum_{k=1}^{\infty}
\sum_{r=0}^{\infty}
\frac{(-1)^{k+r}}{ k}J_{2r+1} 
\left(z_l\right)\\
\cdot
\cos\left((2r+1)(k_l+\varphi)\right),\quad   k_l \coloneqq \lambda\frac{2\pi l}{N},   z_l \coloneqq 2\pi l\frac{A}{\Delta} 
\end{multline}
where the fact that $J_{n}(\cdot)$ is an odd function of its argument when $n$ is odd, has 
been exploited. By multiplying all terms together and by $\frac{16}{N^4}  \cos(k_i-k_u)\cos(k_h-k_l)$, and by summing over 
$i,u,h,l$ we obtain:
\begin{multline}
\frac{16}{N^4}
\sum_{i=0}^{N-1}\sum_{u=0}^{N-1} 
\sum_{h=0}^{N-1}\sum_{l=0}^{N-1}
s_ie_u(s_u)s_he_l(s_l)
\cos\left( k_i-k_u \right) \\
\cdot
\cos\left( k_h-k_l \right)
\\
=\frac{64A^2}{N^4}
\left(\frac{\Delta}{\pi}\right)^2 
\sum_{i=0}^{N-1}
\sum_{u=0}^{N-1} 
\sum_{h=0}^{N-1}
\sum_{l=0}^{N-1}
\sum_{n=1}^{\infty}
\sum_{m=1}^{\infty}
\frac{(-1)^{n+m}}{nm}  \\
\cdot
\sum_{r=0}^{\infty}
\sum_{t=0}^{\infty}
(-1)^{r+t}
J_{2r+1}\left( z_n\right)
J_{2t+1}\left( z_m\right)\\
\cdot
\left\{
\cos\left(k_i+\varphi \right)
\cos\left((2r+1)(k_u+\varphi)\right)
\cos\left(k_h+\varphi \right) \right.\cdot \\
\left.
\cdot
\cos\left((2t+1)(k_l+\varphi)\right)
\right\}
\cos\left( k_i-k_u \right) 
\cos\left( k_h-k_l \right) 
\end{multline}
The  term between curly brackets depends on $\varphi$, so that by taking the 
expectation, we have:
\begin{multline}
E
\left\{
\frac{16}{N^4}
\sum_{i=0}^{N-1}
\sum_{u=0}^{N-1} 
\sum_{h=0}^{N-1}
\sum_{l=0}^{N-1}
s_ie_u(s_u)s_he_l(s_l)
\cos\left( k_i-k_u \right) \right. \\
\cdot
\cos\left( k_h-k_l \right)
\biggr\}
\\
=
\frac{64A^2}{N^4}
\left(\frac{\Delta}{\pi}\right)^2 
\sum_{i=0}^{N-1}
\sum_{u=0}^{N-1} 
\sum_{h=0}^{N-1}
\sum_{l=0}^{N-1}
\sum_{n=1}^{\infty}
\sum_{m=1}^{\infty}
\frac{(-1)^{n+m}}{nm} \\
\cdot
\sum_{r=0}^{\infty}
\sum_{t=0}^{\infty}
(-1)^{r+t}
J_{2r+1}\left( z_n\right)
J_{2t+1}\left( z_m\right)\\
\cdot
E
\left(
C(1,2r+1,1,2t+1) 
\right)
\cos\left( k_i-k_u \right) 
\cos\left( k_h-k_l \right) 
\label{op}
\end{multline}
where the expectation operation has been interchanged with the limit in the infinite series,  because of the 
applicability of the dominated convergence theorem.
Consider that 
the expected value within (\ref{op}) is different from zero only for certain combinations of the indices $r$ and $t$, so that
it behaves like a sieve that  filters some of the terms in the double summation over those indices.
The remaining combinations of indices satisfy the Diophantine equations  
in (\ref{ULHK}) and may lead to simplified versions of
the expressions derived as in (\ref{op}), that are faster to sum numerically. 
The method taken here is applicable to any of the terms in $z_{iuhl}$,
however this last simplification approach becomes more cumbersome for the terms in $z_0$ and in $z_1$. 

Expression (\ref{op}) provides meaningful information when $N\rightarrow \infty$.
In this case, the summations over the indices $i,u,h,l$ can be interchanged to provide:
\begin{multline}
\lim_{N \rightarrow \infty}
E
\left\{
\frac{16}{N^4}
\sum_{i=0}^{N-1}
\sum_{u=0}^{N-1} 
\sum_{h=0}^{N-1}
\sum_{l=0}^{N-1}
s_ie_u(s_u)s_he_l(s_l) \right. \\
\left.
\cdot \cos\left( k_i-k_u \right)
\cos\left( k_h-k_l \right)
\right\}
\\
=
\lim_{N \rightarrow \infty}
\frac{64A^2}{N^4}
\left(\frac{\Delta}{\pi}\right)^2 
\sum_{n=1}^{\infty}
\sum_{m=1}^{\infty}
\frac{(-1)^{n+m}}{nm} \\
\cdot
\sum_{r=0}^{\infty}
\sum_{t=0}^{\infty}
(-1)^{r+t}
J_{2r+1}\left( z_n\right)
J_{2t+1}\left( z_m\right) 
\\
\cdot
\sum_{i=0}^{N-1}
\sum_{u=0}^{N-1} 
\sum_{h=0}^{N-1}
\sum_{l=0}^{N-1}
E
\left(
C(1,2r+1,1,2t+1) 
\right) \\
\cdot
\cos\left( k_i-k_u \right) 
\cos\left( k_h-k_l \right) 
\label{op2}
\end{multline}
Because of (\ref{ULHK}) and (\ref{se}), the rightmost summations in (\ref{op2}) vanish unless
$r=0$ and $t=0$, in which case we have:
\begin{multline}
\lim_{N \rightarrow \infty}
E
\left\{
\frac{16}{N^4}
\sum_{i=0}^{N-1}
\sum_{u=0}^{N-1} 
\sum_{h=0}^{N-1}
\sum_{l=0}^{N-1}
s_ie_u(s_u)s_he_l(s_l) \right.\\
\left.
\cdot
\cos\left( k_i-k_u \right)
\cos\left( k_h-k_l \right)
\right\}
\\
=
4A^2
\left(\frac{\Delta}{\pi}\right)^2 
\left(
\sum_{n=1}^{\infty}
\frac{(-1)^{n}}{n}
J_{1}\left( z_n\right) 
\right)^2
= 4A^2g^2(A,\Delta).
\label{op3}
\end{multline}
This approach applies for any combination of signal and errors in the terms of $z_{iuhl}$, so that when $N\rightarrow \infty$ 
we obtain:
\begin{align}
\begin{split}
z_4 & \rightarrow A^4 \\
z_3 & \rightarrow 8A^3g(A,\Delta) \\
z_2 & \rightarrow  24A^2g^2(A,\Delta),\qquad N \rightarrow \infty\\  
z_1& \rightarrow  32Ag^3(A,\Delta)\\
z_0 & \rightarrow 16g^4(A,\Delta)
\end{split}
\end{align}
and
\begin{multline}
E\left( ({\hat A}^2 )^2\right) \overset{N \rightarrow \infty}{=} A^4+8A^3g(A,\Delta)+24A^2g^2(A,\Delta)\\
+32Ag^3(A,\Delta)+16g^4(A,\Delta),
\end{multline}
that is equal to the square of 
\be
	E({\hat A}^2)  
	\overset{N \rightarrow \infty}{=}
	A^2+4Ag(A,\Delta)+4g^2(A,\Delta),
\ee 
obtained from  (\ref{bias}).
Thus, when $N \rightarrow \infty$ the variance of ${\hat A}^2$ vanishes.

\section{Sum of $g(A,\Delta)$ \label{approxg}}
Because of its role, it is shown next how to sum the series in (\ref{gg}).
Two approaches will be taken.
It can be directly observed that $g(A,\Delta)$ is a 
Schl\"omilch series and its sum is provided by the Nielsen formula as follows 
\cite{Treatise}:
\begin{multline}
g(A,\Delta)=
\frac{\Delta}{\pi}
\sum_{k=1}^{\infty} \frac{(-1)^k}{k}J_1\left( z_k\right)
=\frac{\Delta}{\pi}
\left\{
\vphantom{\sum_{k=1}^p\left[ x^2-\left( k-\frac{1}{2} \right)^2\pi^2\right]^\frac{1}{2}}   
-\frac{x}{2} \right. \\
\left.
+\frac{\sqrt{\pi}}{x\Gamma\left(\frac{3}{2}\right)}
\sum_{k=1}^p\left[ x^2-\left( k-\frac{1}{2} \right)^2\pi^2\right]^\frac{1}{2}
\right\}
\label{schloe}
\end{multline}
{
where $x$ and $p$ are defined in (\ref{pgiven}).}
By observing that $\Gamma\left(\frac{3}{2}\right)=\frac{\sqrt{\pi}}{2}$,
when $p$ is given,
the derivative of (\ref{schloe}) with respect to $A$, is 
\begin{multline}
g'(A,\Delta,p) \coloneqq \frac{\partial g(A,\Delta)}{\partial A} =
-\frac{A}{2}
+\frac{2\Delta}{\pi}\\
\cdot\sum_{k=1}^{p}\frac{\left(k-\frac{1}{2}\right)^2}{A^3\sqrt{1-\frac{\left(k-\frac{1}{2}\right)^2}{A^2}}},\quad
\label{deriv}
\left(p-\frac{1}{2}\right)\Delta \leq A < \left(p+\frac{1}{2}\right)\Delta.
\end{multline}
From (\ref{deriv}) we have:
\[
\lim_{A \rightarrow (p-\frac{1}{2})\Delta^+}  g'(A,\Delta,p)  = +\infty, \qquad p=1, 2, \dots
\]
The numerical sum of (\ref{deriv}), done by assuming $1 \leq p \leq 2^{20}$, shows that 
$g'((p-\frac{1}{2})\Delta,\Delta,p-1)$
converges to $0$ for increasing values of $p$, remaining
always negative. Thus, when $A=\left(p-\frac{1}{2}\right)\Delta$, (\ref{schloe}) is locally minimized.
By substituting these values in (\ref{schloe}), the sequence of the
minima in $g(A,\Delta)$ is: 
\begin{multline}
	g\left( \left(p-\frac{1}{2} \right)\Delta, \Delta\right) = 
	\left(\frac{1}{2}-p\right)\frac{\Delta}{2}\\
+\frac{2\Delta}{\pi}
\sum_{k=1}^{p-1}
\sqrt{1-\frac{\left( k-\frac{1}{2}\right)^2}{\left( p-\frac{1}{2}\right)^2}},
\quad p=0, 1, \ldots
\label{envelope}	
\end{multline}
that represents the discrete envelope of the negative peaks in $g(A,\Delta)$. 
Since when $\Delta \rightarrow 0$, $g^2(A, \Delta) \ll A$, from (\ref{bias}), we 
can write an expression for the sequence of its minima, that is the lower discrete envelope of the 
graphs in Fig.~\ref{figthree},
\begin{align}
\mbox{env}(p,\Delta) &
	\overset{\overset{N \rightarrow \infty }{\Delta \rightarrow 0}}{\simeq}
	4Ag\left( \left(p-\frac{1}{2} \right)\Delta, \Delta\right) \nonumber \\ 
	&\; \; \eqqcolon B_2(\Delta), \quad p=0, 1, \ldots
\label{enve}
\end{align}

Alternatively, the sum in (\ref{schloe}) can be calculated by
using the same approach taken in \cite{Gray}.
Assume $n$ odd. Then, from \cite{TablesIntegrals}
\[
	J_n(x) = \frac{2}{\pi}\int_{0}^{\frac{\pi}{2}} \sin (n\theta) \sin(x \sin \theta) d\theta,
\]
and with $\gamma$ a positive real number,
\begin{align}
\begin{split}
\sum_{k=1}^{\infty} & \frac{(-1)^k}{k} 
 J_{n}(2\pi \gamma k)  =\\
&=\frac{2}{\pi}\int_{0}^{\frac{\pi}{2}} \sin (n\theta) \sum_{k=1}^{\infty} \frac{(-1)^k}{k}
\sin(2\pi k \gamma \sin \theta) d\theta
\\ 
& =
\frac{2}{\pi}\int_{0}^{\frac{\pi}{2}} \sin (n\theta) \left[ \frac{\pi}{2}-\pi \left 
\langle \gamma \sin(\theta) +\frac{1}{2}\right \rangle\right]d\theta
\end{split}
\end{align}
Since
\[
\left \langle \gamma \sin(\theta) +\frac{1}{2}\right \rangle   =
\gamma \sin(\theta) +\frac{1}{2} -\left \lfloor \gamma \sin(\theta) +\frac{1}{2} \right \rfloor
\]
we have
\begin{multline*}
\sum_{k=1}^{\infty} \frac{(-1)^k}{k}
J_{n}(2\pi \gamma k) =
-2\gamma\int_{0}^{\frac{\pi}{2}} \sin (n\theta) 
\sin(\theta)d\theta \\
+2\int_{0}^{\frac{\pi}{2}} \sin (n\theta)\left \lfloor 
\gamma \sin(\theta) +\frac{1}{2}\right \rfloor d\theta
\end{multline*}
The leftmost integral is $0$ unless $n=1$. In this case, it becomes equal to $\frac{\pi}{4}$.
So 
\begin{multline*}
\sum_{k=1}^{\infty} \frac{(-1)^k}{k}
 J_{n}(2\pi \gamma k) =
-\gamma\frac{\pi}{2} \delta_{n-1} \\
+
2\int_{0}^{\frac{\pi}{2}} \sin (n\theta)\left \lfloor 
\gamma \sin(\theta) +\frac{1}{2}\right \rfloor d\theta
\end{multline*}
where $\delta_n$ is equal to $1$ when $n=0$ and $0$ otherwise.
With $k$ integer
\[
\left \lfloor 
\gamma \sin(\theta) +\frac{1}{2}\right \rfloor = k
\implies
k \leq \gamma \sin(\theta) +\frac{1}{2} < k+1
\]
and, assuming $0 \leq \theta \leq \frac{\pi}{2}$, one obtains
\[
\arcsin \frac{k-\frac{1}{2}}{\gamma}
\leq \theta 
<
\arcsin \frac{k+\frac{1}{2}}{\gamma},
\qquad 0 \leq k \leq K
\]
where $K$ is the largest integer $k$ such that $\frac{k+\frac{1}{2}}{\gamma}\leq 1$, that is
$K=\lfloor \gamma -\frac{1}{2}\rfloor$. 
When $\gamma-\frac{1}{2} \neq$ integer number define 
\[
b_k \coloneqq
\left\{
\begin{array}{ll}
\arcsin \frac{k-\frac{1}{2}}{\gamma} & 0 < k \leq K \\ 
0 & k=0 \\
1 & k = K+1
\end{array}
\right.
\]
Conversely, if $\gamma-\frac{1}{2}$ is an integer number, define
\[
b_k \coloneqq
\left\{
\begin{array}{ll}
\arcsin \frac{k-\frac{1}{2}}{\gamma} & 0 < k \leq K \\ 
0 & k=0 \\
\end{array}
\right.
\]
Then, for odd $n$, 
\begin{multline}
 \sum_{k=1}^{\infty} \frac{(-1)^k}{k}
 J_{n}(2\pi \gamma k) =
-\gamma\frac{\pi}{2} \delta_{n-1} \\
+
2\int_{0}^{\frac{\pi}{2}} \sin (n\theta)\left \lfloor 
\gamma \sin(\theta) +\frac{1}{2}\right \rfloor d\theta \\ =
-\gamma\frac{\pi}{2} \delta_{n-1}
-\frac{2}{n}
\sum_{k=0}^{\overline{K}} k \left[ \cos(n b_{k+1})-\cos(n b_k)\right]
\label{sumgray}
\end{multline}
where
$\overline{K}$ is equal to $K$ if $\gamma-\frac{1}{2} $ is not integer and to $K+1$, conversely.
By assuming $n=1$ in (\ref{sumgray}), an alternative expression for $g(A,\Delta)$ can easily be otained.

Neither (\ref{schloe}) nor (\ref{sumgray}) 
provide direct information on the rate of convergence of $g(A,\Delta)$ to $0$ when $\Delta$
goes to $0$, as expected.
A somewhat loose upper bound can be obtained by considering that
\cite{landau}:
\[
|J_1(x)| \leq \frac{c}{|x|^{1/3}},
\]
where $c=0.7857\ldots$.
Consequently, when $A>0$,
{
\[
|g(A,\Delta)| \leq  \frac{\Delta}{\pi}
\sum_{k=1}^{\infty}   
\frac{1}{k}  
\frac{c}{\left( \frac{2\pi k A}{\Delta}\right)^{\frac{1}{3}}}
= B(A,\Delta),
\]
where $B(A,\Delta)$ is defined in (\ref{badelta}).}
Therefore, under the assumption that $N \rightarrow \infty$,
\be
|bias(A,\Delta)| \leq  4AB(A,\Delta)+4B(A,\Delta)^2
\eqqcolon B_1(A,\Delta),
\label{B1}
\ee
and  $bias(A,\Delta)  
\sim O\left( \Delta^{\frac{4}{3}}\right)$
when $\Delta  \rightarrow 0$, for a given value of $A$.
Observe also that $B_1(A,\Delta)$ is minimum when $A=\frac{1}{2}$.

%

\end{document}